\documentclass[a4paper,extra,final,onecolumn]{gji}  % referee

\usepackage{timet}  % redefines font
\usepackage[utf8]{inputenc} %unicode support
\usepackage{url}
\usepackage{graphicx}
\usepackage{booktabs}
\usepackage{multirow}
\usepackage{amsmath}
    \DeclareMathOperator{\arctantwo}{arctan2}
\usepackage{amssymb}
\usepackage{array}
\usepackage{dsfont}

\usepackage{siunitx}
    \DeclareSIUnit{\nT}{\nano\tesla}
    \DeclareSIUnit{\yr}{yr}
    \DeclareSIUnit{\sfu}{sfu}
\usepackage{gji-siunitx-patch}
\usepackage{mathtools}
    \DeclarePairedDelimiter{\abs}{\lvert}{\rvert}%

\usepackage{upgreek}
\let\bibhang\relax
\let\bibfont\relax
\usepackage[%
	backend=biber,%
	bibencoding=utf8,
	natbib=false,%
	style=authoryear-comp,%
	citestyle=authoryear-comp,%
    dashed=false,%  % replace duplicate authors in the bib with a dash
	maxbibnames=5,%
	maxcitenames=1,%
	uniquelist=false,%
	uniquename=false,%
	sorting=nyt,%  % name, year, title
	url=false,%
    date=year,%  % only show year, not month + year
	doi=true,%
	eprint=false,%
	hyperref=true,%
	backref=false,%
]{biblatex}
\usepackage{footnote}
    \makesavenoteenv{tabular}
    \makesavenoteenv{table}
\usepackage[pdftex,linktoc=all,colorlinks=true,linkcolor=blue,citecolor=blue,pagebackref=false,hyperfootnotes=false, bookmarksopen=true]{hyperref}

\date{Accepted 2023 August 13. Received 2023 August 10; in original form 2023 March 8}
\pagerange{\pageref{firstpage}--\pageref{lastpage}}
\volume{00}
\pubyear{2023}

\title{Polar ionospheric currents and high temporal resolution geomagnetic field models}
\author[C.\ Kloss, C.\ C.\ Finlay, K.\ M.\ Laundal, and N.\ Olsen]{%
  Clemens Kloss$^1$, Christopher C.\ Finlay$^1$, Karl M.\ Laundal$^2$, and Nils Olsen$^1$ \\
  $^1$ DTU Space, Technical University of Denmark, Centrifugevej 356, 2800 Kongens Lyngby, Denmark \\
  $^2$ Birkeland Centre for Space Science, Department of Physics and Technology, University of Bergen, 5007 Bergen, Norway
}%

\newcommand{\swarm}{\textit{Swarm}}
\newcommand{\modela}{\mbox{\textit{Model-A}}}
\newcommand{\modelb}{\mbox{\textit{Model-B}}}
\newcommand{\modelref}{\mbox{\textit{Reference}}}

\newcommand{\tabitem}[1]{~~\llap{#1}~~}

\begin{document}

\label{firstpage}

\maketitle

\begin{summary}
    Estimating high resolution models of the Earth's core magnetic field and its time variation in the polar regions requires that one can adequately account for magnetic signals produced by polar ionospheric currents, which vary on a wide range of time and length scales. Limitations of existing ionospheric field models in the challenging polar regions can adversely affect core field models, which in turn has important implications for studies of the core flow dynamics in those regions. Here we implement a new approach to co-estimate a climatological model of the ionospheric field together with a model of the internal and magnetospheric fields within the CHAOS geomagnetic field modelling framework. The parametrization of the ionospheric field exploits non-orthogonal magnetic coordinates to efficiently account for the geometry of the Earth's magnetic field and scales linearly with external driving parameters related to the solar wind and the interplanetary magnetic field. Using this approach we derive a new geomagnetic field model from measurements of the magnetic field collected by low Earth orbit satellites, which in addition to the internal field provides estimates of the typical current system in the polar ionosphere and successfully accounts for previously unmodelled ionospheric signals in field model residuals. To resolve the ambiguity between the internal and ionospheric fields when using satellite data alone, we impose regularisation. We find that the time derivative of the estimated internal field is less contaminated by the polar currents, which is mostly visible in the zonal and near-zonal terms at high spherical harmonic degrees. Distinctive patches of strong secular variation at the core-mantle boundary, which have important implications for core dynamics, persist. Relaxing the temporal regularisation reveals annual oscillations, which could indicate remaining ionospheric field or related induced signals in the internal field model. Using principal component analysis we find that the annual oscillations mostly affect the zonal low-degree spherical harmonics of the internal field.

    % Better to remove reference to resolution here because it is rather loosely used here. We don't investigate the formal resolution in the paper, so this could be misleading. I rather had leakage in mind although resolution is certainly also affected.
    % Too strong to say that the results are sensitive to the model regularization. This statement casts doubt on all the obtained results. It a given that we need regularisation and that it affects the results. Be more specific about what is affected, all results? Or rephrase: "We impose regularisation such that ..."
    % Not only degree 2 are affected by annual oscillations. We don't want that this statement triggers people to look for sources that have this specific degree.
    % Future work should focus on alternative ways to accomplish the separation: Rephrase or be more specific on what is meant by alternative ways. Otherwise this is better placed in the discussion or conclusions. Abstract is about what you find.
\end{summary}

\begin{keywords}
    Core, Magnetic field variations through time, Satellite magnetics, Inverse theory, Polar ionospheric currents
\end{keywords}

\section{Introduction}

The ionospheric magnetic field is generated by electrical currents that circulate in the Earth's ionosphere, the electrically conducting layer of the atmosphere from about \SI{90}{\kilo\meter} to \SI{1000}{\kilo\meter} altitude. The ionospheric field undergoes daily, seasonal and solar cycle variations, which depend on solar activity and illumination \parencite{Yamazaki2016}. In particular in the polar regions, where the ionospheric field is highly dynamic and very sensitive to changes in the solar wind and the Interplanetary Magnetic Field (IMF) thanks to field-aligned currents that facilitate the coupling to the magnetosphere, it is a difficult task to estimate accurate ionospheric field models. Imperfect modelling and the fact that the involved time and length scales overlap with those of the large-scale time-varying internal field, which originates in the Earth's core, makes the separation between the two fields a major challenge in geomagnetic field modelling \parencite{Finlay2016}. New strategies to deal with the ionospheric field are therefore crucial for studies of the core field and its time variation. Such core field models are used to infer flow patterns in Earth's core and to study the geodynamo process and its related waves and oscillations. The high latitude regions of the outer core shell, within what is known as the inner core tangent cylinder (a cylinder aligned with the rotation axis just touching the inner core in the equatorial plane), play a special role in core processes because they are dynamically separated from the remainder of the shell and so can be an important source of equatorial symmetry breaking. Recent studies indicate strong jet like flows near to the tangent cylinder \parencite{Livermore2017}, while highly time-dependent turbulent polar vortices are expected within the tangent cylinder \parencite{Aurnou2003,Schaeffer2017,Sheyko2018}. Detailed study of such processes requires that ionospheric signals in the polar regions are adequately separated.

Earlier studies have developed different techniques to model the ionospheric field or have tried to reduce its effect on the recovered internal field during geomagnetic field modelling. One common technique to minimise the ionospheric disturbance field in geomagnetic field modelling is data selection. By focusing on data under geomagnetic quiet conditions and by choosing suitable magnetic components, one seeks to reduce ionospheric signals that are not well parametrized in the model. For example, the CHAOS model \parencite{Olsen2006,Olsen2009,Olsen2010,Olsen2014,Finlay2020}, which is a model of the recent geomagnetic field and provides estimates of the time-dependent and static internal fields and the quiet-time magnetospheric field, is derived from magnetic vector and total intensity observations using, among other criteria, a dark-time selection criterion based on the sun elevation angle to remove the strong ionospheric disturbances that are present under sunlit conditions. At polar latitudes only the scalar magnitude of the field is used in an effort to minimise the effect of field-aligned currents, which mainly disturb the field direction but not its scalar magnitude. Similarly, in the sequential approach of \cite{Ropp2020} for modelling the internal, quiet-time magnetospheric and associated internally-induced fields, the magnetic vector data at mid and low latitudes are selected according to local night-time and dark conditions, but in the polar regions vector data from all local times are used. Although data selection is effective, significant ionospheric signals often remain in the data, especially at polar latitudes, where strong horizontal currents in the E-layer of the ionosphere continue to disturb the scalar magnitude of the field at all local times including during dark conditions \parencite[e.g.][]{FriisChristensen2017}.

In the comprehensive modelling approach \parencite{Sabaka2020,Sabaka2018,Sabaka2015,Sabaka2004,Sabaka2002}, all major sources of the geomagnetic field are parametrized and co-estimated in a single step, including the magnetic field produced by ionospheric currents. In the CM6 model \parencite{Sabaka2020}, the latest in the series of Comprehensive Models (CM), the magnetic field due to the currents in the E-layer of the ionosphere are parametrized in space using special basis functions that involve projecting spherical harmonics in the quasi-dipole coordinate system \parencite{Richmond1995}, which is believed appropriate for describing these currents, whose geometry is organised by the main magnetic field. Temporal variations are expressed in terms of specific daily and sub-daily harmonics with periods of \SI{24}{\hour}, \SI{12}{\hour}, \SI{8}{\hour}, and \SI{6}{\hour}, which are further modulated with annual and semiannual harmonics and scaled by a three-monthly average of $F_{10.7}$ solar radiation index, which tracks long term variations in solar activity. In addition, the model takes into account the Earth-induced field using a model of the electrical conductivity of the Earth's surface. Thanks to the sophisticated parametrization, the CM models are very successful at describing the slowly varying averaged ionospheric magnetic field at mid and low latitudes. This is why the same parametrization has been adopted in the dedicated ionospheric field inversion chain \parencite{Chulliat2013} to produce spherical harmonic models of the ionospheric magnetic field at low-to-mid latitudes using the magnetic data collected by the satellites of the European Space Agency's (ESA) \swarm{} mission \parencite{FriisChristensen2006}. However, the approach of using a finite set of specific harmonics may not be as suitable for describing the polar ionospheric field, which varies on a much wider range of frequencies in response to changes in the solar wind speed and the IMF. In addition, it is not clear whether the basis functions for parametrizing the ionospheric magnetic field are also appropriate at high latitudes.

In the Kalmag geomagnetic field models \parencite{Baerenzung2020,Baerenzung2022} the magnetic field associated with ionospheric currents including field-aligned currents are also co-estimated. These models are sequentially derived using a Kalman filter approach after applying data selection to reduce the dayside ionospheric field signal in the input magnetic data. For the parametrization of the ionospheric sources they use poloidal and toroidal potentials in magnetic coordinate systems and represent the evolution in time through random processes based on a-priori spatio-temporal statistics. \cite{Lesur2008} estimate ionospheric currents in the polar ionosphere as part of the first generation of the GFZ Reference Internal Magnetic Models (GRIMM). However, they did not co-estimate the ionospheric field but had to use a two-step procedure whereby they first derived the model part corresponding to the internal, large-scale external and associated induced fields from the data and then used its residuals to build the ionospheric part of the model.

Apart from geomagnetic field models, there are dedicated ionospheric field models such as the Average Magnetic field and Polar current System (AMPS) model \parencite{Laundal2018} that seek to better model the fields and currents in the polar regions. Instead of using specific periodicities to model the variability of the ionospheric disturbance field explicitly in time as in the CM models, the AMPS model focuses on its climatological aspects, i.e.\ it seeks to model the long term average of the field as a function of external driving parameters. The AMPS model expresses the ionospheric field in terms of poloidal and toroidal potentials, which are expanded into a global basis of spherical harmonics. It exploits magnetic apex coordinates \parencite{Richmond1995} to efficiently take into account the geometry of the main magnetic field, which organises the large-scale spatial structure of the ionospheric field. To express the variability of the average ionospheric field in time, the model uses a combination of external driving parameters related to the solar wind speed and IMF components. It is, however, derived using vector residuals, i.e.~observations of the magnetic vector taken by the CHAllenging Minisatellite Payload (CHAMP) and \swarm{} satellites after the removal of estimates of the internal and magnetospheric fields given by the CHAOS model.

In this study we combine the climatological approach of the AMPS model for modelling the ionospheric field with the CHAOS framework for modelling the internal and magnetospheric fields. More specifically, we implement a co-estimation approach, where an AMPS-type ionospheric field model is derived at the same time as a geomagnetic field model similar to the CHAOS model. Making use of satellite magnetic observations made by the CHAMP, CryoSat-2 and \swarm{} satellites during geomagnetic quiet conditions, we derive a new model of the geomagnetic field. Using this model, we study the quiet-time ionospheric field and the associated electrical currents in the polar regions and go on to investigate the effect on the time-variation of the internal field at polar latitudes when ionospheric currents are co-estimated. In addition, we explore cases when the temporal smoothness imposed on the internal field model is considerably relaxed. Note that the goal is not to derive an all-purpose model of the ionospheric field but rather to improve the time-dependent internal field model in the CHAOS modelling framework for geomagnetic quiet conditions in the challenging polar regions.

The paper is organised as follows. In Sect.~\ref{sec:magnetic_observations_and_data_selection} we describe the satellite magnetic data used and the applied data selection. In Sect.~\ref{sec:model_parameterization_and_estimation} we provide details about the model parametrization, giving special attention to the ionospheric part taken from the AMPS model. There, we also give the equations for the model estimation and the applied regularisation, and list the chosen regularisation parameters. In Sect.~\ref{sec:results} we evaluate the performance of the estimated model in terms of the fit to the magnetic data, study the polar ionospheric currents during geomagnetic quiet conditions and investigate the recovered core field and its time variations at polar latitudes. In the last part of that section we study the variations in time of the internal field as given by a test model where we apply weaker temporal smoothing. We finish with a discussion of the obtained results in Sect.~\ref{sec:discussion} and the conclusions in Sect.~\ref{sec:conclusions}.

\section{Magnetic observations and data selection}
\label{sec:magnetic_observations_and_data_selection}

We used vector observations of the magnetic field made by the CHAMP and CryoSat2 satellites, and the three satellites of the \swarm{} constellation, Swarm-A, Swarm-B and Swarm-C, from 2001 to the end of 2021.

From the CHAMP mission, we used the Level 3 \SI{1}{\hertz} magnetic data, product CH-ME-3-MAG \parencite[]{Rother2019}, between January 2001 and August 2010, which we downsampled to \SI{1}{\minute} values. We selected data according to the recommended quality flags that are provided in the distributed CHAMP data product files \parencite[]{GFZSection2019}. However, we did not require that both star camera heads on the boom close to the vector magnetometer were active and provided attitude information at the time of measurement since this created gaps in the global distribution of magnetic data at low and mid latitudes during dusk and dawn. More specifically, we allowed data if at least one of the two star camera heads was available. To account for the corresponding increase in the uncertainty of the attitude information, we chose larger a-priori attitude errors compared to when both star camera heads were active (see Sect.~\ref{sec:data_error_covariances} for details).

Concerning CryoSat2, we used fully calibrated \SI{4}{\second} magnetic vector data from the onboard fluxgate magnetometer FGM1 (CryoSat2-1), version 0103, from August 2010 to the end of 2013. These data have been calibrated as described in \cite{Olsen2020}. We reduced the dataset to \SI{1}{\minute} values through the following steps. First, we used estimates of the time-dependent internal field and the CryoSat2-1 Euler angles from CHAOS-7.9 to compute residuals in the calibrated magnetometer frame. Then, we performed a Huber-weighted linear regression of the residuals within \SI{20}{\second} intervals and kept one fit value from each interval. Finally, we added back the previously subtracted model estimates but retained only every third value to obtain a reduced time series of approximately \SI{1}{\minute} resolution. By using \SI{20}{\second} intervals for the linear fit, we followed \cite{Olsen2020}, who recommends averaging over five successive values to reduce the intrinsic noise. In addition, we removed data if the attitude uncertainty $q_\mathrm{error}$, which is provided in the CryoSat2 data product files, was larger than \SI{40}{arcseconds}.

From the \swarm{} mission, we made use of the Level 1b \SI{1}{\hertz} magnetic vector data from all three satellites (Swarm-A, Swarm-B and Swarm-C), versions 0505-0508 as available, from November 2013 to the end of 2021. We downsampled the magnetic data from each satellite to \SI{3}{\minute} values to have a similar amount of data per time interval as for CHAMP and CryoSat2.

On the entire dataset of magnetic observations, we applied several selection criteria to focus on geomagnetic quiet-time conditions. First, we removed gross outliers for which vector residuals with respect to the CHAOS-7.9 field model were greater than \SI{1000}{\nT}. We note that this approach also removed magnetic signals in the data associated with field-aligned currents, which can reach several thousands of \si{\nT} in the polar regions also during geomagnetic quiet-time conditions. Similarly, the averaging of the CryoSat-2 data, as described above, removed high-frequency ionospheric magnetic signals, i.e., signals that varied along the satellite orbit on timescales much shorter than the 20-second interval used for averaging. Nevertheless, since we do not expect that our approach of modelling the average ionospheric field is able to capture intermittent high-amplitude events, we preferred to remove these data and to process the CryoSat-2 data in this way to improve the overall quality of the model. Next, to focus on geomagnetically quiet conditions, we selected data for which the $\mathit{Kp}$ index \parencite[]{Matzka2021,Matzka2021a} was below 2o and the absolute rate of change of the $\mathit{RC}$ index \parencite[]{Olsen2014}, a quantitative measure of the magnetic disturbance at equatorial and mid-latitudes similar to the $\mathit{Dst}$ index \parencite[]{Sugiura1991}, was below \SI{2}{\nT\per\hour}. Furthermore, we selected data if, on average over \SI{2}{\hour} prior to the time of measurement, the Newell coupling function [\cite[][for the exact definition used in this study, see Eq.~\eqref{eq:coupling_Newell}]{Newell2007}], measuring the rate of magnetic flux opened at the magnetopause, was below \num{2.4} and the IMF at the magnetopause was pointing northward, i.e.\ having a positive z-component in the Geocentric Solar Magnetic (GSM) frame.

\begin{figure*}
    \includegraphics{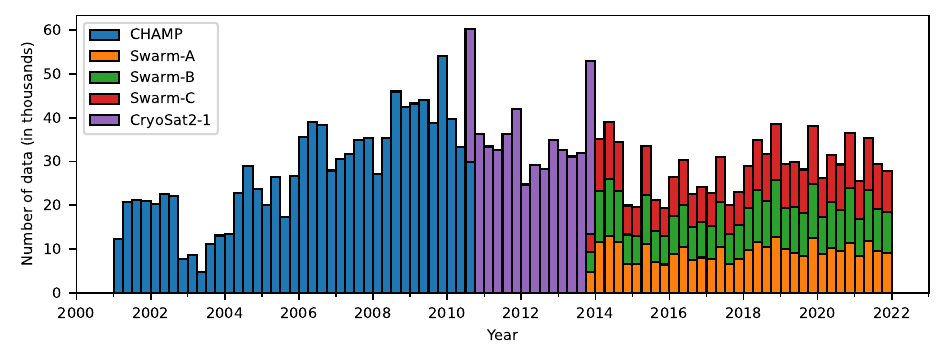}
    \caption{Number of selected vector data every 3 months for each satellite shown as stacked histogram.}
    \label{fig:data_distribution}
\end{figure*}

The data processing and selection resulted in $N_d = \num{2472746}$ vector observations, which we used for estimating models of the geomagnetic field. To illustrate the data distribution in time, we show in Fig.~\ref{fig:data_distribution} a stacked histogram of the amount of data in 3-month intervals for each satellite dataset.
We did not treat data differently depending on dark and sunlit conditions during the model estimation to avoid seasonal variations in the data distribution. Otherwise using, for example, only dark data for the estimation of the internal field would adversely affect the time-dependence of the associated model parameters, unless sufficiently smoothed in time through regularisation. This is due to an annual variation in the data distribution, which is created by the periodic exclusion of the data in the polar region on the summer hemisphere. Similarly, we did not select data based on magnetic local time for the estimation of the internal field to uniformly sample the polar electrojets. Note that we did not use ground-based magnetic field observations as input data for the modelling both to allow comparisons of the model predictions with independent data and to be sure not to bias the geographical distribution of the input data.

\section{Model parametrization and estimation}
\label{sec:model_parameterization_and_estimation}

In this paper, we largely follow the modelling approach of the CHAOS geomagnetic field model series \parencite[]{Olsen2006,Olsen2009,Olsen2010,Olsen2014}, version CHAOS-7.9 \parencite[]{Finlay2020}. However, a significant difference to the CHAOS models is that we also co-estimate a model of the ionospheric currents based on the AMPS model \parencite[]{Laundal2018}. The following summarises the parametrization of the geomagnetic field sources that are represented in our model and gives the equations used for the model parameter estimation.

\subsection{Internal magnetic field}

Satellites in low Earth orbit take magnetic measurements in a region that is free of electrical currents associated with the internal sources. In the quasi-static approximation, the internal magnetic field can therefore be represented by an internal scalar potential, $V^\mathrm{int}$, such that $\mathbf{B}^\mathrm{int} = -\nabla V^\mathrm{int}$. In spherical coordinates $V^\mathrm{int}$ is given by
\begin{equation}
    V^\mathrm{int}(t, r, \theta, \phi) = a\sum_{n=1}^{N^\mathrm{int}}\sum_{m=-n}^{n}\bigg(\frac{a}{r}\bigg)^{n+1} g_n^m(t)Y_n^m(\theta, \phi),
\end{equation}
where $a=\SI{6371.2}{\kilo\meter}$ is the mean surface radius of the Earth, $g_n^m$ are the spherical harmonic coefficients of degree $n$ and order $m$, $Y_n^m$ are the spherical harmonic functions, and $N^\mathrm{int}=55$ is the chosen truncation degree to limit the spatial resolution of the model. The spherical harmonic functions are defined as
\begin{equation}
    Y_n^m(\theta, \phi) \equiv \left\{
        \begin{aligned}
             & \cos(m\phi) P_n^m(\cos\theta), \, m\geq 0 \\
             & \sin(\abs{m}\phi) P_n^{\abs{m}}(\cos\theta), \, m < 0,
        \end{aligned}
    \right.
\end{equation}
where $\theta$ and $\phi$ are respectively the geocentric colatitude and longitude, and $P_n^m$ are the associated Legendre functions using the Schmidt semi-normalization. We allow the spherical harmonic coefficients for $n \leq 20$ to be time-dependent using a basis of 6th-order B-splines to account for the slow time changes of the internal field
\begin{equation}
    g_n^m(t) = \sum_{k=1}^{K} g_{n,k}^m \mathcal{B}_{6, k}(t),
\end{equation}
where $\mathcal{B}_{6, k}$ ($k=1, \dots, K$) are the B-spline basis functions defined on the model interval using a sequence of knots with a \SI{0.5}{\yr} knot spacing and a 6-fold knot multiplicity at the model endpoints. The coefficients for $21 \leq n \leq N^\mathrm{int}$ are kept constant to represent the high-degree part of the assumed static lithospheric field.

\subsection{External magnetic field}

The sources of the external field are located in the space above the Earth's surface. In our model we distinguish between the magnetospheric field and the ionospheric field. The parametrization of the magnetospheric field is identical to the CHAOS model, whereas the one for the ionospheric field is basically taken from the AMPS model. In this section, we will also introduce magnetic apex coordinate systems, which are important for an efficient parametrization of the ionospheric magnetic field.

\subsubsection{Ionospheric field}

Following the approach of \cite{Laundal2016} we write the ionospheric magnetic field as
\begin{equation}
    \label{eq:ionospheric_field}
    \mathbf{B}^\mathrm{ion} = \mathbf{B}^\mathrm{pol} + \mathbf{B}^\mathrm{tor} = -\nabla V^\mathrm{ion} + \hat{\mathbf{r}}\times\nabla T^\mathrm{ion},
\end{equation}
where the poloidal magnetic field $\mathbf{B}^\mathrm{pol}$, written in terms of the scalar potential $V^\mathrm{ion}$, is associated with the currents in the ionospheric E-layer, which flow entirely below the measurement shell of the satellites, whereas the toroidal magnetic field $\mathbf{B}^\mathrm{tor}$, written in terms of the potential $T^\mathrm{ion}$, is associated with the field-aligned currents that couple the polar ionosphere to the magnetosphere (Birkeland currents). To take advantage of the fact that the currents in the ionosphere are highly organised with respect to the geomagnetic field, we specify the potentials in magnetic apex coordinate systems defined by \cite{Richmond1995}. There are two of these systems: Quasi-Dipole (QD) and Modified-Apex (MA). In QD coordinates the latitude is defined as
\begin{equation}
    \lambda_\mathrm{QD} = \pm \arccos\sqrt{\frac{a+h}{a + h_\mathrm{A}}},
\end{equation}
where positive (negative) values refer to the northern (southern) magnetic hemisphere, $h$ is the geodetic height of the point of interest, and $h_\mathrm{A}$ is the geodetic height of the apex, which is the highest point above the Earth's ellipsoidal surface along the magnetic field line, as given by a geomagnetic field model, that passes through the point of interest. The longitude of the QD coordinate system is defined as the longitude of the apex in centred dipole coordinates, a coordinate system where the z-axis points along Earth's magnetic dipole axis towards the northern hemisphere, the y-axis is perpendicular to both the dipole axis and the rotation axis, and the x-axis completes the right-handed system \parencite[]{Laundal2017}. In Modified-Apex (MA) coordinates the latitude is defined as
\begin{equation}
    \lambda_\mathrm{MA} = \pm \arccos\sqrt{\frac{a+h_\mathrm{R}}{a + h_\mathrm{A}}},
\end{equation}
where $h_\mathrm{R}$ is a chosen reference height for the mapping, which we set to $h_\mathrm{R} = \SI{110}{\kilo\meter}$. The MA latitude is positive for points that map to the northern magnetic hemisphere and negative otherwise. The longitude of the MA coordinate system is identical to the QD longitude. Since both are equal, they can be used interchangeably. Coordinates and base vectors of the two magnetic apex coordinate systems can be conveniently computed with the Python software package Apexpy \parencite[]{Meeren2021}, which is a wrapper of the Fortran library by \cite{Emmert2010}. As the reference model for the field line tracing, we used the 13th generation of the International Geomagnetic Reference Field \parencite[IGRF;][]{Alken2021} at epoch 2015.0 throughout the entire model time interval.

Using a combination of the apex coordinate systems, again following \cite{Laundal2016}, we express the ionospheric potentials in terms of spherical harmonic functions
\begin{subequations}
    \begin{align}
        V^\mathrm{ion}(h, \theta_\mathrm{QD}, \phi_\mathrm{MLT}) &= a\sum_{n=1}^{N^\mathrm{ion}}
        \sum_{\substack{m=-n \\ \abs{m}\leq M}}^n
        \left(\frac{a}{a+h}\right)^{n+1}g_n^{m, \mathrm{ion}}Y_n^m(\theta_\mathrm{QD},
        \phi_\mathrm{MLT}) \label{eq:ionospheric_field_potential_a}\\
        T^\mathrm{ion}(\theta_\mathrm{MA}, \phi_\mathrm{MLT}) &= (a+h_\mathrm{R})\sum_{n=1}^{N^\mathrm{tor}}
        \sum_{\substack{m=-n \\ \abs{m}\leq M}}^n T_n^{m,\mathrm{ion}} Y_n^m(\theta_\mathrm{MA},
        \phi_\mathrm{MLT}),
    \end{align}
\end{subequations}
where $\theta_\mathrm{QD} = \frac{\uppi}{2} - \lambda_\mathrm{QD}$ and $\theta_\mathrm{MA} = \frac{\uppi}{2} - \lambda_\mathrm{MA}$ are the QD and MA colatitudes, respectively. We chose to truncate the spherical harmonic representations at $N^\mathrm{ion} = 45$ and
$N^\mathrm{tor}=65$. In addition, we used a maximum spherical harmonic order of $M = 3$ for both potentials in agreement with \cite{Laundal2018}. Instead of the QD and MA longitudes, we used the Magnetic Local Time (MLT)
\begin{equation}
    \phi_\mathrm{MLT} = \phi_\mathrm{QD} - \phi_\mathrm{noon} + \uppi,
\end{equation}%
where $\phi_\mathrm{noon}$ is the QD longitude of the subsolar point, computed on a sphere with radius $r \gg a $, in practice $r = 50a$. Using MLT takes account of the fact that the ionospheric field stays fixed with respect to the sun. By writing $T^\mathrm{ion}$ only in dependence of the MA latitude and MLT, we assume the potential to be constant along the IGRF magnetic field lines; we do not however impose north-south symmetry.

By inserting QD colatitude and MLT into the spherical harmonic functions in Eq.~\ref{eq:ionospheric_field_potential_a}, we assume that $V^{ion}$ defines a harmonic potential in the source-free region. To test this assumption, we performed numerical computations and found that $-\nabla V^\mathrm{ion}$ is approximately but not strictly divergence-free. The deviations from zero, which are largest in the auroral regions and along the magnetic dip equator, are usually smaller in absolute value than \SI{1}{\nT}. This is smaller than typical errors due to other unmodelled sources, which remain larger in the polar regions despite co-estimating a climatological model of the ionospheric magnetic field. Our conclusions from these tests is therefore that $V^\mathrm{ion}$, organised in magnetic apex coordinates and magnetic local time, satisfactorily approximates a potential field and is useful for parametrizing the geometry of the ionospheric magnetic field and current densities.

Inserting the expressions for the potentials into Eq.~\eqref{eq:ionospheric_field} and evaluating the gradients yields
\begin{subequations}
    \begin{align}
        \mathbf{B}^\mathrm{pol} &= -\frac{1}{(a + h)\sin\theta_\mathrm{QD}}\frac{\partial
        V^\mathrm{ion}}{\partial \phi_\mathrm{MLT}} \mathbf{f}_2\times\hat{\mathbf{k}} - \frac{1}{a + h}\frac{\partial
        V^\mathrm{ion}}{\partial \theta_\mathrm{QD}} \mathbf{f}_1\times\hat{\mathbf{k}} -
        \sqrt{\abs{\mathbf{f}_1\times\mathbf{f}_2}}\frac{\partial V^\mathrm{ion}}{\partial h}\hat{\mathbf{k}} \label{eq:ionospheric_field_long_a}\\
        \mathbf{B}^\mathrm{tor} &=
        \frac{1}{(a + h_\mathrm{R})\sin\theta_\mathrm{MA}}\frac{\partial T^\mathrm{ion}}{\partial
        \phi_\mathrm{MLT}}\hat{\mathbf{k}}\times\mathbf{d}_1 +
        \frac{\sqrt{4-3\sin^2\theta_\mathrm{MA}}}{2(a + h_\mathrm{R})\cos\theta_\mathrm{MA}}\frac{\partial T^\mathrm{ion}}{\partial
        \theta_\mathrm{MA}}\hat{\mathbf{k}}\times\mathbf{d}_2
    \end{align}
\end{subequations}
where $\{\mathbf{d}_1,\mathbf{d}_2, \mathbf{f}_1, \mathbf{f}_2\}$ are the non-orthogonal base vectors for the magnetic apex coordinate systems \parencite[]{Laundal2017}, and $\hat{\mathbf{k}}$ is a unit vector in the geodetic upward direction. Here, we used that $\hat{\mathbf{r}}\approx\hat{\mathbf{k}}$, which means that the expressions are best suited for describing the ionospheric field in the polar regions. Nevertheless we assume that they also approximate the ionospheric field at low latitudes well. Note that the last term in Eq.~\eqref{eq:ionospheric_field_long_a} is multiplied with $\sqrt{\abs{\mathbf{f}_1\times\mathbf{f}_2}}$ based on the assumption that the vertical component of the ionospheric field scales with the linear dimension of the horizontal current system \parencite[]{Richmond1995}.

The ionospheric magnetic field can be related to an electric sheet current density (in units of \si{\ampere\per\meter}) that flows at a fixed height, chosen to be $h_\mathrm{R}$, written in the form of
\begin{equation}
    \label{eq:sheet_current}
    \mathbf{J}^\mathrm{sh} = \mathbf{J}^\mathrm{df} + \mathbf{J}^\mathrm{cf} = \hat{\mathbf{k}}\times\nabla \psi^\mathrm{df} + \nabla \psi^\mathrm{cf},
\end{equation}%
where $\mathbf{J}^\mathrm{df}$ is the divergence-free part of the sheet current density associated with $\mathbf{B}^\mathrm{pol}$ and $\mathbf{J}^\mathrm{cf}$ is the curl-free part associated with $\mathbf{B}^\mathrm{tor}$. The potentials of the sheet current density parts are
\begin{subequations}
    \label{eq:sheet_current_potentials}
    \begin{align}
        \psi^\mathrm{df}(t, \theta_\mathrm{QD}, \phi_\mathrm{MLT}) &= -\frac{a}{\mu_0}\sum_{n=1}^{N^\mathrm{ion}} \sum_{\substack{m=-n \\ \abs{m}\leq M}}^n\frac{2n+1}{n}\left(\frac{a}{a+h_\mathrm{R}}\right)^{n+1} g_n^{m, \mathrm{ion}}(t) Y_n^m(\theta_\mathrm{QD}, \phi_\mathrm{MLT}) \label{eq:sheet_current_potentials_a}\\
        \psi^\mathrm{cf}(t, \theta_\mathrm{MA}, \phi_\mathrm{MLT}) &= -\frac{a + h_\mathrm{R}}{\mu_0}\sum_{n=1}^{N^\mathrm{tor}}\sum_{\substack{m=-n \\ \abs{m}\leq M}}^n T_n^{m, \mathrm{ion}}(t) Y_n^m(\theta_\mathrm{MA}, \phi_\mathrm{MLT}), \label{eq:sheet_current_potentials_b}
    \end{align}
\end{subequations}%
which were derived by treating the apex coordinates as if they were orthogonal, following the approach of \cite{Laundal2018}. The curl-free part of the sheet current density can be furthermore related to an upward current density $J_u$ (in units of \si{\ampere\per\meter\squared}) through $J_u = -\nabla\cdot\mathbf{J}^\mathrm{cf}$, a statement of current continuity, which yields at the reference height
\begin{equation}
    J_u(t, \theta_\mathrm{MA}, \phi_\mathrm{MLT}) = -\frac{1}{\mu_0 (a + h_\mathrm{R})}\sum_{n=1}^{N^\mathrm{tor}}\sum_{\substack{m=-n \\ \abs{m}\leq M}}^n n(n+1) T_n^{m, \mathrm{ion}}(t) Y_n^m(\theta_\mathrm{MA}, \phi_\mathrm{MLT}).
\end{equation}%
At polar latitudes, where the magnetic field lines are close to vertical, $J_u$ can be interpreted as field-aligned currents and $\mathbf{J}^\mathrm{cf}$ as the horizontal closure of these currents in the form of a sheet current.

\begin{figure*}
    \includegraphics{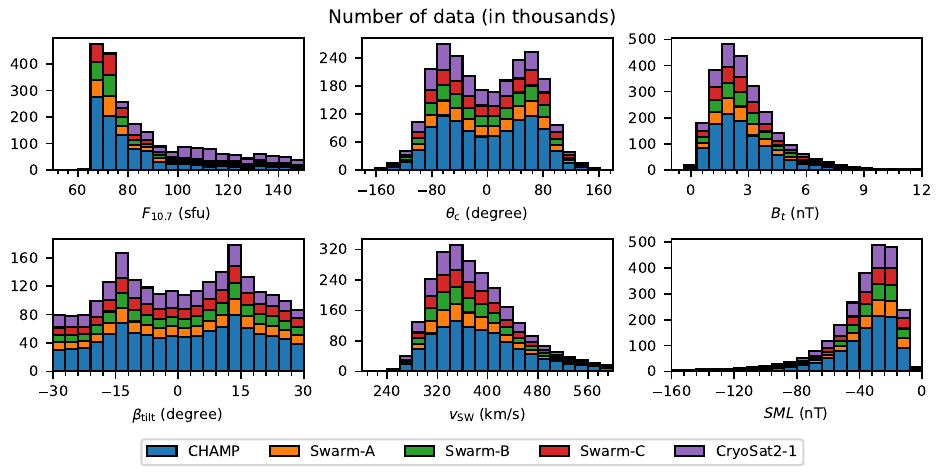}
    \caption{Stacked histograms showing the number of selected magnetic vector data in dependence of the driving parameters for the ionospheric field at the time of measurement. The colours indicate the different satellite data sets.}
    \label{fig:external_parameter_distribution}
\end{figure*}

Instead of parametrizing the expansion coefficients $g_n^{m, \mathrm{ion}}$ and $T_n^{m, \mathrm{ion}}$ in time explicitly, we followed the climatological approach of the AMPS model \parencite[]{Laundal2018} and wrote these coefficients as linear combinations of external driving parameters $X_i$ ($i=1,\dots,19$) so that
\begin{equation}
    g_n^{m,\mathrm{ion}}(t) = g_{n, 0}^{m, \mathrm{ion}} + \sum_{i=1}^{19} g_{n, i}^{m, \mathrm{ion}} X_i(t)
\end{equation}
and similarly for $T_n^{m,\mathrm{ion}}$. The $X_i$ are combinations of solar wind parameters and IMF components that have been found suitable for characterising the external driving of the ionospheric current system \parencite[]{Laundal2018,Weimer2013}
\begin{equation}
    \label{eq:external_parameters}
    \begin{aligned}
        &X_1 = \sin\theta_\mathrm{c} & &X_2 = \cos\theta_\mathrm{c} & &X_3 = \epsilon & &X_4 = \epsilon\sin\theta_\mathrm{c} \\
        &X_5 = \epsilon\cos\theta_\mathrm{c} & &X_6 = \beta_\mathrm{tilt} & &X_7 = \beta_\mathrm{tilt}\sin\theta_\mathrm{c} & &X_8 = \beta_\mathrm{tilt}\cos\theta_\mathrm{c}\\
        &X_{9} = \epsilon\beta_\mathrm{tilt} & &X_{10} = \epsilon \beta_\mathrm{tilt}\sin\theta_\mathrm{c} & &X_{11} =\epsilon\beta_\mathrm{tilt}\cos\theta_\mathrm{c} & &X_{12} =\tau \\
        &X_{13} =\tau\sin\theta_\mathrm{c} & &X_{14} =\tau\cos\theta_\mathrm{c} & &X_{15} =\tau\beta_\mathrm{tilt} & &X_{16} =\tau\beta_\mathrm{tilt}\sin\theta_\mathrm{c} \\ &X_{17} =\tau\beta_\mathrm{tilt}\cos\theta_\mathrm{c} & &X_{18} = F_{10.7} & &X_{19} =\mathit{SML}, & &
    \end{aligned}
\end{equation}
which are all functions of time. The terms in Eq.~\eqref{eq:external_parameters} involve the clock angle
\begin{equation}
    \theta_\mathrm{c} = \arctantwo(B_{\mathrm{IMF},y}, B_{\mathrm{IMF},z}),
\end{equation}
where the components of the IMF, $B_{\mathrm{IMF},y}$ and $B_{\mathrm{IMF},z}$, are with respect to the GSM frame; the dipole tilt angle
\begin{equation}
    \beta_\mathrm{tilt}= \arcsin(\hat{\mathbf{s}}\cdot\hat{\mathbf{m}}_\mathrm{dip}),
\end{equation}
where $\hat{\mathbf{s}}$ is a unit vector in the direction of the sun and $\hat{\mathbf{m}}_\mathrm{dip}$ is the dipole moment of the IGRF magnetic field, parametrizes seasonal effects; the solar wind-magnetospheric coupling function \parencite[]{Newell2007}
\begin{equation}
    \label{eq:coupling_Newell}
    \epsilon = \num{e-3}\abs{v_\mathrm{sw}}^{4/3}{B_t}^{2/3}
    \sin^{8/3}\frac{\abs{\theta_\mathrm{c}}}{2},
\end{equation}
where $B_t = \sqrt{B^2_{\mathrm{IMF},y} + B^2_{\mathrm{IMF},z}}$ (given in \si{nT}) and $v_\mathrm{sw}$ (given in \si{\kilo\meter\per\second}) is the solar wind velocity component antiparallel to the x-axis of the GSM frame, maximises for southward IMF and measures the rate of reconnection on the dayside magnetopause; and the coupling function
\begin{equation}
    \tau = \num{e-3}\abs{v_\mathrm{sw}}^{4/3}{B_t}^{2/3}
    \cos^{8/3}\frac{\theta_\mathrm{c}}{2}
\end{equation}
maximises for northward IMF and measures the rate of lobe reconnection in the magnetotail. To approximate the delay of the near-Earth space environment to adjust to changes in the external driving, we used \SI{20}{\minute} moving averages of $\epsilon$, $\tau$ and the clock angle, based on \SI{1}{\minute} values propagated to the magnetopause as provided by the OMNI database \parencite[]{King2005}. The solar radiation index $F_{10.7}$ in units of solar flux, $\si{\sfu} \equiv \SI{e-22}{\watt\per\meter\squared\per\hertz}$, parametrizes solar cycle variations. Expanding on the original parametrization of AMPS, we included as $X_{19}$ the $\mathit{SML}$ index \parencite[]{Newell2011}, developed by the SuperMAG initiative \parencite[]{Gjerloev2009,Gjerloev2012}, to parametrize indirectly driven currents in the polar ionosphere. With typically more than 100 contributing ground-based magnetometer stations, the $\mathit{SML}$ index can be considered an extension of the traditionally used $\mathit{AL}$ index, which is based on only 12 stations, to monitor nightside auroral activity. Fig.~\ref{fig:external_parameter_distribution} shows stacked histograms of the number of selected magnetic vector data in dependence of the external driving parameters at the time of measurement.

\subsubsection{Magnetospheric field}

The magnetic field $\mathbf{B}^\mathrm{mag}$ produced by electric currents in the magnetosphere can be separated into contributions due to the ring current in the near-Earth magnetosphere, $\mathbf{B}^\mathrm{near}$, and the currents in the remote magnetosphere, $\mathbf{B}^\mathrm{far}$, so that $\mathbf{B}^\mathrm{mag} = \mathbf{B}^\mathrm{near} + \mathbf{B}^\mathrm{far}$. We present the parametrization of each contribution in detail below.

The magnetic field produced by the ring current in the near-Earth magnetosphere is written as $\mathbf{B}^\mathrm{near} = -\nabla V^\mathrm{near}$ using a scalar potential in Solar Magnetic (SM) coordinates, where the z-axis is anti-parallel to the dipole axis of the  Earth's magnetic field, the x-axis is in the plane spanned by the dipole axis and the Earth-Sun line, and the y-axis completes the right-handed system \parencite[]{Laundal2017}. The scalar potential is given by
\begin{equation}
    \begin{aligned}
        V^\mathrm{near}(t, r, \theta_\mathrm{SM}, \phi_\mathrm{SM}) &= a\sum_{n=1}^{N^\mathrm{near}}\sum_{m=-n}^{n} \left(\frac{r}{a}\right)^n q_n^{m, \mathrm{SM}}(t) Y_n^m(\theta_\mathrm{SM}, \phi_\mathrm{SM}) \\
        &+ a\sum_{m=-1}^{1} \hat{q}_1^{m, \mathrm{SM}}\bigg[\mathit{RC}_i(t)\left(\frac{a}{r}\right)^2 + \mathit{RC}_e(t)\left(\frac{r}{a}\right)\bigg]Y_1^m(\theta_\mathrm{SM}, \phi_\mathrm{SM}) \\
        &+ \text{Earth-induced counterpart}
    \end{aligned}
\end{equation}
where $N^\mathrm{near} = 2$ is the chosen truncation degree, $\hat{q}_1^{m, \mathrm{SM}}$ are constant regression parameters multiplying the $\mathit{RC}$ index, which consists of an internal part, $\mathit{RC}_i$, and an external part, $\mathit{RC}_e$, so that $\mathit{RC} = \mathit{RC}_i + \mathit{RC}_e$. We estimated the spherical harmonic coefficients $q_n^{m, \mathrm{SM}}$ with $n=1$, called $RC$-baseline corrections, in bins of 30 days, except for a single bin covering the period from August 2010 to January 2014, when only platform magnetometer data of CryoSat2 was available. The coefficients for $n=2$ were treated as constants over the entire model time interval. The potential of the internally induced field was not estimated separately but coupled to the external potential by means of Q-responses, which are based on models of Earth's electrical conductivity. These Q-responses have also been used for the decomposition of the $\mathit{RC}$ index into internal and external parts. The reader is referred to \cite{Finlay2020} for details concerning the treatment of induced fields in CHAOS-7, which is also the approach used here.

The magnetic field produced by the remote sources in the magnetosphere, assumed to primarily be the magnetopause and magnetotail currents, is written as $\mathbf{B}^\mathrm{far} = -\nabla V^\mathrm{far}$ using an axisymmetric, static scalar potential in GSM coordinates, where the x-axis points sunward along the Earth-Sun line, the z-axis is contained within the dipole axis and the Earth-Sun line, while the y-axis completes the right-handed system \parencite[]{Laundal2017}. The potential is given by
\begin{equation}
    \begin{aligned}
        V^\mathrm{far}(r, \theta_\mathrm{GSM}, \phi_\mathrm{GSM}) &= a\sum_{n=1}^{N^\mathrm{far}} \left(\frac{r}{a}\right)^n q_n^{0, \mathrm{GSM}} Y_n^0(\theta_\mathrm{GSM}, \phi_\mathrm{GSM}) \\
        &+ \text{Earth-induced counterpart}
    \end{aligned}
\end{equation}
where $N^\mathrm{mag}=2$ is the chosen truncation degree. The treatment of the internally induced part is similar to the approach used for  $\mathbf{B}^\mathrm{near}$.

\subsection{Alignment parameters}

We estimate alignment parameters in the form of three Euler angles $\alpha$, $\beta$ and $\gamma$ for each satellite to rotate the magnetic vector components from the frame of the Vector Field Magnetometer (VFM) to the Common Reference Frame (CRF), which is defined by the orientation of the onboard star cameras. The alignment can be written in matrix notation as
\begin{equation}
\mathbf{B}_\mathrm{CRF} = \mathbf{R}_3(\gamma) \mathbf{R}_2(\beta) \mathbf{R}_1(\alpha) \mathbf{B}_\mathrm{VFM},
\end{equation}
where $\mathbf{B}_\mathrm{CRF}$ and $\mathbf{B}_\mathrm{VFM}$ are column vectors that contain the magnetic field components with respect to the CRF and the VFM frame, respectively, and $\mathbf{R}_1$, $\mathbf{R}_2$ and $\mathbf{R}_3$ are rotation matrices given by
\begin{equation}
    \mathbf{R}_1(\alpha) =
    \begin{pmatrix}
    1 & 0 & 0 \\
    0 & \cos\alpha & -\sin\alpha \\
    0 & \sin\alpha & \cos\alpha \\
    \end{pmatrix},
    \quad
    \mathbf{R}_2(\beta) =
    \begin{pmatrix}
    \cos\beta & 0 & \sin\beta \\
    0 & 1 & 0 \\
    -\sin\beta & 0 & \cos\beta \\
    \end{pmatrix},
    \quad
    \mathbf{R}_3(\gamma) =
    \begin{pmatrix}
    \cos\gamma & -\sin\gamma & 0 \\
    \sin\gamma & \cos\gamma & 0 \\
    0 & 0 & 1 \\
    \end{pmatrix}.
\end{equation}
Another rotation based on the quaternions provided in the data product files, which describe the rotation from CRF to an Earth-fixed frame in dependence on satellite position and orientation, was then performed to obtain the vector magnetic field in terms of geocentric spherical components. We estimated the above Euler angles in bins of 30 days to allow for time variations.

\subsection{Model estimation}

The model parameters are arranged into a column vector
$\mathbf{m} = [\mathbf{p}^\mathrm{T},\mathbf{q}^\mathrm{T}]^\mathrm{T}$, where the column vector $\mathbf{p}$ contains the parameters of the geomagnetic field model and the column vector $\mathbf{q}$ contains the Euler angles for the alignment of the magnetic vector observations. We solved for the model parameter vector by iteratively minimising the following cost function using a quasi-Newton scheme
\begin{equation}
    \Phi(\mathbf{m}) = [\mathbf{g}(\mathbf{p}) - \mathbf{d}(\mathbf{q})]^\mathrm{T}\mathbf{C}_d^{-1}[\mathbf{g}(\mathbf{p}) - \mathbf{d}(\mathbf{q})] + \mathbf{m}^\mathrm{T}\boldsymbol{\Lambda}\mathbf{m}
\end{equation}
where $\mathbf{g}(\mathbf{p})$ is a column vector containing the model estimates of the magnetic vector components, $\mathbf{d}(\mathbf{q})$ is a column vector containing the aligned magnetic vector observations expressed in terms of spherical geocentric components, $\mathbf{C}_d^{-1}$ is the inverse of the data error covariance matrix, and $\boldsymbol{\Lambda}$ is the model regularisation matrix. At each iteration $k$ the model parameter vector is updated through
\begin{equation}
    \mathbf{m}_{k+1} = \mathbf{m}_k +  (\mathbf{G}_k^\mathrm{T}\mathbf{C}_\mathrm{d}^{-1}\mathbf{G}_k + \mathbf{\Lambda})^{-1} [\mathbf{G}_k^\mathrm{T}\mathbf{C}_\mathrm{d}^{-1}(\mathbf{d}_k-\mathbf{g}_k) - \mathbf{\Lambda}\mathbf{m}_k],
\end{equation}
where $\mathbf{g}_k = \mathbf{g}(\mathbf{p}_k)$, $\mathbf{d}_k = \mathbf{d}(\mathbf{q}_k)$ and $\mathbf{G}_k$ is the matrix of partial derivatives of the residuals with respect to the model parameter vector
\begin{equation}
    \big(\mathbf{G}_k\big)_{ij} = \frac{\partial [\mathbf{g}(\mathbf{p})-\mathbf{d}(\mathbf{q})]_i}{\partial (\mathbf{m})_j}\bigg|_{\mathbf{m} = \mathbf{m}_k}.
\end{equation}

\subsubsection{Data error covariances}
\label{sec:data_error_covariances}

In the data error covariance matrix we account for the instrument error and the uncertainty in the attitude information provided by the star trackers. The error contributions are most conveniently described in the B23 frame, which is defined by unit base vectors in the direction of $\mathbf{B}$, $\hat{\mathbf{n}}\times\mathbf{B}$ and $\mathbf{B}\times(\hat{\mathbf{n}}\times\mathbf{B})$, where $\hat{\mathbf{n}}$ is the star camera bore sight assumed not parallel to $\mathbf{B}$. In this reference frame the data error covariance matrix is diagonal and given by \cite{Holme1996}
\begin{equation}
    \mathbf{C}_\mathrm{B23} = \mathrm{diag}[\sigma^2, \sigma^2 + B^2\chi^2 - (\chi^2 - \psi^2)(\hat{\mathbf{n}}\cdot\mathbf{B})^2, \sigma^2 + B^2\psi^2]
\end{equation}
where $\sigma$ (in \si{\nT}) is an isotropic instrument error in the vector component magnitudes, $\chi$ (in radians) is an error in the attitude about $\hat{\mathbf{n}}$, $\psi$ (in radians) is an error in the attitude about the two axes perpendicular to $\hat{\mathbf{n}}$. In the B23 frame, we multiplied the inverse of the data error covariance matrix, which is diagonal, by the Huber weights \parencite[]{Huber2004,Constable1988}, which we recomputed from the residuals at each iteration. The use of Huber weights allows the robust estimation of model parameters in the presence of long-tailed error distributions due to sources of error besides the instrument and attitude errors. We also applied a $\sin\theta$ weighting to compensate for the larger amount of data near the poles due to the high-inclination orbits of the satellites. Tab.~\ref{tab:attitude_error} gives an overview of the used a-priori instrument and attitude errors, which are based on results of previous modelling efforts, most notably the CHAOS model series.
\begin{table*}
    \centering
    \caption{Adopted instrument and attitude errors for the satellite datasets.}
    \label{tab:attitude_error}
    \begin{tabular}{lrrr}
        \toprule
         &  $\sigma$ (\si{\nT}) & $\psi$ (arcsec) & $\chi$ (arcsec) \\
        Dataset &  &  &  \\
        \midrule
        CHAMP & 2.5 & 10 & 10$^\dagger$ \\
        CryoSat2-1 & 6.0 & 30 & 30 \\
        Swarm-A & 2.2 & 5 & 5 \\
        Swarm-B & 2.2 & 5 & 5 \\
        Swarm-C & 2.2 & 5 & 5 \\
        \bottomrule
        \multicolumn{4}{l}{%
            \raisebox{-0.5em}{\parbox[t]{7cm}{$^{\dagger}$ \footnotesize When both head units of the star camera are active, otherwise \SI{60}{arcseconds}.}}
        }
    \end{tabular}
\end{table*}
The star camera bore sight $\hat{\mathbf{n}}$, which is aligned with the z-axis of the CRF frame, was taken from the data product files. Note that star cameras often consist of several head units, in which case the bore sight direction is a weighted average of the directions of the individual head units that were active at the time of measurement. Also note that in our case, the value of $\hat{\mathbf{n}}$ is in fact arbitrary since we assume $\psi$ and $\chi$ to be equal. However, $\hat{\mathbf{n}}$ is important in the case of CHAMP when only one of the two head units was active at a time.

\subsubsection{Model regularisation}

The model regularisation matrix $\boldsymbol{\Lambda}$ aids the convergence of the model estimation by applying smoothing penalties on the model parameters. It is a block-diagonal matrix, where each block corresponds to a penalty measure scaled with an adjustable parameter, the regularisation parameter. To reduce the temporal variation of the internal field, we used a regularisation term based on the squared value of the third time-derivative of the radial internal field at the Core-Mantle Boundary (CMB) ($r = \SI{3485}{\kilo\meter}$), averaged over both the entire model time interval and the CMB, and another regularisation term based on the squared value of the second time-derivative of the radial internal field at the CMB, evaluated at the model endpoints, $t_\mathrm{s} = 2001.0$ and $t_\mathrm{e} = 2022.0$ in units of decimal years, and averaged over the CMB. The corresponding regularisation parameters are $\lambda_t$, $\lambda_{t_s}$ and $\lambda_{t_e}$, respectively. The time variation of each $\mathit{RC}$-baseline correction, $\{q_{1}^{-1, \mathrm{SM}}, q_{1}^{0, \mathrm{SM}}, q_{1}^{1, \mathrm{SM}}\}$, was minimised using a quadratic norm of the bin-to-bin differences, which is scaled by the regularisation parameter $\lambda_\mathrm{mag}$. For the ionospheric field, we implemented two regularisation terms. For the first term, instead of directly applying a regularisation on the poloidal ionospheric magnetic field $\mathbf{B}^\mathrm{pol}$, we designed a quadratic norm based on the associated divergence-free sheet currents in the ionospheric E-layer. More specifically, this regularisation term is based on the squared magnitude of the average divergence-free sheet currents as seen by an Earth-fixed observer, integrated over the spherical surface, which can be written as a quadratic form
\begin{equation}
    \mathbf{m}^\mathrm{T}\boldsymbol{\Lambda}^\mathrm{pol}\mathbf{m} = \sum_{s\in\{r, \theta, \phi\}} \frac{1}{4\uppi} \int_{S(r_0)}\left[\frac{1}{N_d}\sum_{i=1}^{N_d} J_s^\mathrm{df}(t_i, r_0, \theta, \phi) \right]^2 \sin\theta \mathrm{d}\theta \mathrm{d}\phi,
\end{equation}
where $S(r_0)$ is the spherical surface of radius $r = r_0 \equiv a+h_\mathrm{R}$, $J_s^\mathrm{df}$ with $s \in \{r, \theta, \phi\}$ are the geocentric spherical components of the divergence-free sheet current density [see Eqs.~\eqref{eq:sheet_current} and \eqref{eq:sheet_current_potentials_a}] as given by the model. The surface integral was implemented by, first, computing the components of the divergence-free sheet current density on a Gauss-Legendre grid in spherical geocentric coordinates given the external driving parameter values at the times $t_i$ in the input dataset, then, forming the arithmetic mean of each component, and, finally, integrating the sum of the squared component means over the sphere using the integration weights. When strongly enforced by choosing a large value of the associated regularisation parameter $\lambda_\mathrm{pol}$, the regularisation pushes to zero the component means of the divergence-free sheet current density with respect to an Earth-fixed frame. Note that the currents can still change in time as required by the magnetic data but only to the extent that the time average remains small. We found that this form of regularisation helps to resolve the ambiguity between the internal field and the poloidal ionospheric field, which is caused by the fact that both fields have sources that are internal with respect to the satellites. Without the regularisation, the internal field showed artefacts in the form of near-zonal patterns that were almost time-invariant and parallel to lines of constant QD latitude and the divergence-free sheet currents were organised into a single cell of current encircling the magnetic poles, very different from the expected configuration consisting of two cells of current separated by the noon-midnight meridian. Regarding the second regularisation term, for the toroidal ionospheric field, we followed the AMPS model by using a regularisation term based on the spatial power spectrum of the toroidal field to prevent large amplitudes close to the magnetic dip equator, where the mapping of points at satellite altitude to MA coordinates leaves a gap in $\theta_\mathrm{MA}$. The associated regularisation matrix $\boldsymbol{\Lambda}^\mathrm{tor}$ is diagonal with entries $\frac{n(n+1)}{2n+1}$, which depend on the degree of the expansion coefficients $T_n^{m, \mathrm{ion}}$. The associated regularisation parameter is $\lambda_\mathrm{tor}$.

We derived three geomagnetic field models: the first model, referred to as \modela{}, accounts for the ionospheric field and is our preferred model, whereas the second model, denoted \modelref{}, is identical except for omitting the ionospheric part. The third model, \modelb{} is identical to \modela{}, but we reduced the temporal regularisation of the internal field part of the model. Tab.~\ref{tab:summary} summarises the parametrization of the three models and gives the numerical values of the regularisation parameters used in this study.
{%
\renewcommand\baselinestretch{1.45}  % prevent from overly stretching table
\begin{table}
    \centering
    \caption{Summary of the model parametrization and the chosen numerical values of the regularisation parameters for the three estimated models. The description of the parametrization is further divided into spatial (S) and temporal (T), when applicable.}
    \label{tab:summary}
    \begin{tabular}{l>{\centering\arraybackslash}p{3.4cm}>{\centering\arraybackslash}p{3.4cm}>{\centering\arraybackslash}p{3.4cm}}
        \toprule
        \textbf{Internal field} & & & \\[\defaultaddspace]
        Time-dependent field & \multicolumn{3}{l}{\tabitem{S:}Spherical harmonics in geographic coordinates ($n \leq 20$)}\\
         & \multicolumn{3}{l}{\tabitem{T:}6th-order B-splines, \SI{0.5}{\yr} knot spacing, 6-fold endpoint multiplicity}\\
        Static field & \multicolumn{3}{l}{\tabitem{S:}Spherical harmonics in geographic coordinates ($21 \leq n \leq 55$)} \\
         & \multicolumn{3}{l}{\tabitem{T:}Static in geographic coordinates} \\
        \midrule
        \textbf{Ionospheric field} & & & \\[\defaultaddspace]
        & Reference & \multicolumn{2}{c}{Model-A/Model-B}\\
        \cmidrule(l{1.5ex}r{1.5ex}){1-4}
        Poloidal field & n/a & \multicolumn{2}{l}{\tabitem{S:}Spherical harmonics in QD/MLT,} \\
         & & \multicolumn{2}{l}{\hphantom{\tabitem{S:}}$n \leq 45$, $m \leq 3$} \\
         & & \multicolumn{2}{l}{\tabitem{T:}19 external driving parameters + constant} \\
        Toroidal field & n/a & \multicolumn{2}{l}{\tabitem{S:}Spherical harmonics in MA/MLT,} \\
         & & \multicolumn{2}{l}{\hphantom{\tabitem{S:}}$n \leq 65$, $m \leq 3$} \\
         & & \multicolumn{2}{l}{\tabitem{T:}19 external driving parameters + constant} \\
        \midrule
        \textbf{Magnetospheric field} & & &\\[\defaultaddspace]
        Near-magnetospheric field & \multicolumn{3}{l}{\tabitem{S:}Spherical harmonics in SM, $n \leq 2$} \\
         & \multicolumn{3}{l}{\tabitem{T:}Degree-1 spherical harmonic coefficients scaled by hourly $\mathit{RC}$ index,} \\
         & \multicolumn{3}{l}{\hphantom{\tabitem{T:}}degree-2 coefficients static in SM, $\mathit{RC}$-baseline corrections estimated} \\
         & \multicolumn{3}{l}{\hphantom{\tabitem{T:}}in bins of 30 days, except for a single bin between 2010-08/2014-01} \\
        Far-magnetospheric field & \multicolumn{3}{l}{\tabitem{S:}Spherical harmonics in GSM, $n \leq 2$, $m = 0$} \\
         & \multicolumn{3}{l}{\tabitem{T:}Static in GSM} \\
        \midrule
        \textbf{Alignment} & & & \\[\defaultaddspace]
        CHAMP & \multicolumn{3}{l}{3 Euler angles estimated in 118 bins of 30 days length} \\
        Swarm-A & \multicolumn{3}{l}{3 Euler angles estimated in 99 bins of 30 days length} \\
        Swarm-B & \multicolumn{3}{l}{3 Euler angles estimated in 99 bins of 30 days length} \\
        Swarm-C & \multicolumn{3}{l}{3 Euler angles estimated in 99 bins of 30 days length} \\
        CryoSat2-1 & \multicolumn{3}{l}{3 Euler angles estimated in 42 bins of 30 days length} \\
        \midrule
        \textbf{Regularisation} & & & \\[\defaultaddspace]
         & Reference & Model-A & Model-B \\
        \cmidrule(l{1.5ex}r{1.5ex}){1-4}
        $\lambda_t$ (\si{[\nT\per\yr\cubed]^{-2}}) & \num{1.0} & \num{1.0} & \num{0.0125} \\
        $\lambda_{t_s}$ (\si{[\nT\per\yr\squared]^{-2}}) & \num{e-2} & \num{e-2} & \num{1.25e-4} \\
        $\lambda_{t_e}$ (\si{[\nT\per\yr\squared]^{-2}}) & \num{e-2} & \num{e-2} & \num{1.25e-4} \\
        $\lambda_\mathrm{mag}$ (\si{[\nT\per\yr]^{-2}}) & \num{5e3} & \num{5e3} & \num{5e3} \\
        $\lambda_\mathrm{pol}$ (\si{[\milli\ampere\per\meter]^{-2}}) & n/a & \num{e6} & \num{e6} \\
        $\lambda_\mathrm{tor}$ (\si{\nT^{-2}}) & n/a & \num{e5} & \num{e5} \\
        \bottomrule
    \end{tabular}
\end{table}
}%

The iterative minimisation for the model estimation was initialised with a starting model $\mathbf{m}_0$, which we chose to be CHAOS-7.9 for the internal and magnetospheric model parts and, if included, zero-valued parameters for the ionospheric model part. Concerning the Euler angles, we used the initial values that have been determined during the pre-flight calibration on ground for CHAMP \parencite[]{Schwintzer2002} and Swarm \parencite[]{ToeffnerClausen2019}, and during in-flight calibration for CryoSat2-1 \parencite[]{Olsen2020}. Convergence was typically reached after 15 iterations.

\section{Results}
\label{sec:results}

In the following, we report on the achieved misfit of \modelref{} and \modela{}, present the estimated ionospheric field of \modela{}, and compare the estimated internal fields of \modelref{} and \modela{}. \modelb{} is presented in the second half of this section, where we study the effect of relaxing the temporal regularisation of the internal field model when an ionospheric field is co-estimated.

\subsection{Fit to the magnetic data}

\begin{figure*}
    \centering
    \includegraphics{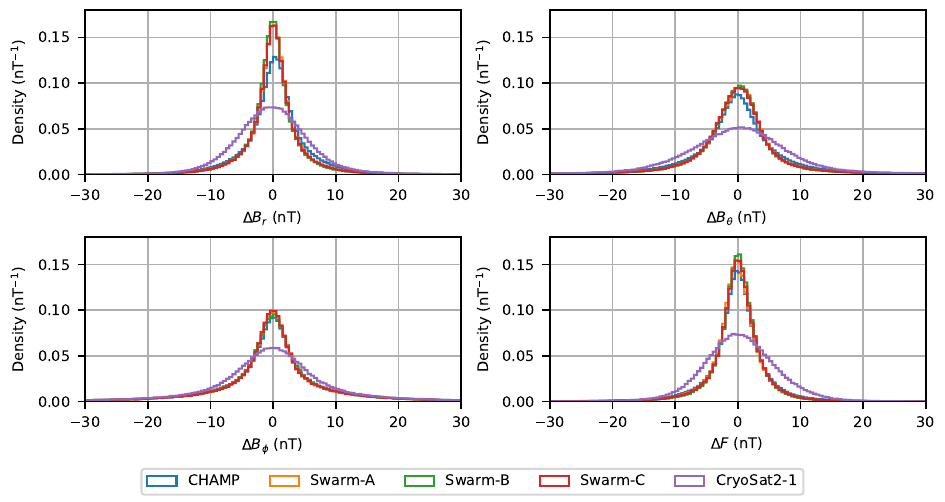}
    \caption{Histograms of vector and scalar residuals for each satellite dataset with respect to \modela{}. The bin width is \SI{0.5}{\nT} and the histograms are normalised to integrate to unit area. The residuals outside the range of \SI{\pm 30}{\nT} are not shown but taken into account for the normalisation.}
    \label{fig:data_residual_histograms}
\end{figure*}

To illustrate how well the magnetic field estimates of \modela{} fit the magnetic data, we computed vector and scalar residuals, i.e.\ the component-wise differences $\Delta B_r$, $\Delta B_\theta$ and $\Delta B_\phi$, and the difference in scalar magnitude, $\Delta F$, between the vector estimates of the magnetic field from \modela{} and the magnetic observations in the dataset used for the model estimation. Note that although the scalar component was not used in constructing the models, it is a useful diagnostic and included here. Fig.~\ref{fig:data_residual_histograms} presents histograms of the vector and scalar residuals for each satellite in bins of \SI{0.5}{\nT} width.
Irrespective of the residual component and the satellite, the histograms are fairly symmetric and have a single maximum close to zero. The peaks of the histograms for the three \swarm{} satellites are narrow and very similar in appearance to the extent that they practically overlap. The peaks for CHAMP are slightly broader with correspondingly lower maxima, especially regarding the radial and southward components. The histograms for CryoSat2-1 are even broader with maxima that are at approximately half the value of the \swarm{} and CHAMP satellites, reflecting the generally higher noise level of these data. \modela{} clearly fits the radial and scalar components better than the $\theta$ and $\phi$ components, which are more influenced by rapidly varying field-aligned currents, which our parametrization does not capture.

\begin{figure*}
    \centering
    \includegraphics{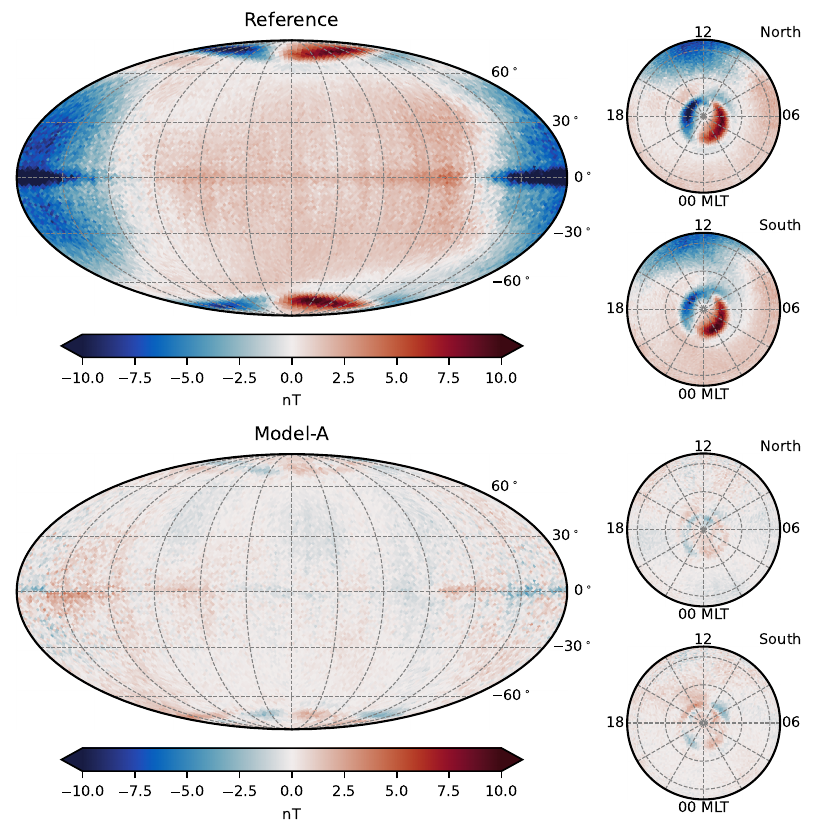}
    \caption{Median scalar residuals with respect to the reference model (top) and \modela{} (bottom) in QD/MLT coordinates using an equal-area pixelation. The global maps on the left are Mollweide projections, where the central vertical line corresponds to midnight ($\phi_\mathrm{MLT}=\SI{0}{\degree}$) and the central horizontal line to the magnetic dip equator ($\lambda_\mathrm{QD}=\SI{0}{\degree}$). The maps on the right are orthographic projections of the northern magnetic hemisphere (North) and, as if looking down on the Earth from the north pole, the southern magnetic hemisphere (South). The labels indicate noon (12), midnight (00), dawn (06) and dusk (18). The dashed lines show parallels and meridians at \SI{30}{\degree} intervals.}
    \label{fig:scalar_residual_maps}
\end{figure*}

Tab.~\ref{tab:misfit} presents Huber-weighted mean and Root-Mean-Square (RMS) values for each satellite and field component, distinguishing between polar ($\abs{\lambda_\mathrm{QD}} > \SI{55}{\degree}$) and non-polar ($\abs{\lambda_\mathrm{QD}} \leq \SI{55}{\degree}$) latitudes.
{%
\renewcommand\baselinestretch{1.15}  % prevent from overly stretching table
\begin{table}
    \centering
    \caption{Statistics of vector and scalar residuals with respect to \modela{}
    and \modelref{} for each satellite dataset. In the table $N$ is the
    number of data, and Mean and RMS refer to the Huber-weighted mean and
    the Huber-weighted root-mean-square of the deviation from the mean,
    respectively. Polar QD latitude refers to $\abs{\lambda_\mathrm{QD}}
    > \SI{55}{\degree}$ and non-polar QD latitude to the opposite.}
    \label{tab:misfit}
    \begin{tabular}{lllrrrrr}
        \toprule
         &  &  & $N$ & \multicolumn{2}{c}{Model-A} & \multicolumn{2}{c}{Reference} \\ \cmidrule(lr){5-6}\cmidrule(lr){7-8}
         &  &  &  & Mean (\si{\nT}) & RMS (\si{\nT}) & Mean (\si{\nT}) & RMS (\si{\nT}) \\
        Dataset & QD latitude & Residual &  &  &  &  &  \\
        \midrule
        \multirow[c]{8}{*}{CHAMP} & \multirow[c]{4}{*}{Non-polar} & $\Delta B_r$ & 661357 & 0.43 & 4.18 & 0.38 & 5.27 \\
         &  & $\Delta B_{\theta}$ & 661357 & -0.13 & 4.62 & -0.05 & 4.94 \\
         &  & $\Delta B_{\phi}$ & 661357 & 0.05 & 5.92 & 0.07 & 6.48 \\
         &  & $\Delta F$ & 661357 & -0.03 & 2.97 & -0.23 & 3.96 \\
        \cmidrule{2-8}
         & \multirow[c]{4}{*}{Polar} & $\Delta B_r$ & 410810 & 0.14 & 6.60 & 0.25 & 8.70 \\
         &  & $\Delta B_{\theta}$ & 410810 & -0.01 & 17.35 & 0.15 & 21.07 \\
         &  & $\Delta B_{\phi}$ & 410810 & 0.03 & 18.89 & 0.31 & 23.14 \\
         &  & $\Delta F$ & 410810 & -0.11 & 6.20 & -0.75 & 8.52 \\
        \cmidrule{1-8}
        \multirow[c]{8}{*}{CryoSat2-1} & \multirow[c]{4}{*}{Non-polar} & $\Delta B_r$ & 285278 & 0.02 & 4.97 & 0.03 & 5.46 \\
         &  & $\Delta B_{\theta}$ & 285278 & -0.40 & 6.16 & -0.43 & 6.23 \\
         &  & $\Delta B_{\phi}$ & 285278 & -0.02 & 6.32 & 0.11 & 6.66 \\
         &  & $\Delta F$ & 285278 & 0.39 & 4.82 & 0.35 & 4.80 \\
        \cmidrule{2-8}
         & \multirow[c]{4}{*}{Polar} & $\Delta B_r$ & 178099 & -0.61 & 7.02 & -0.57 & 7.67 \\
         &  & $\Delta B_{\theta}$ & 178099 & 1.01 & 18.18 & 0.94 & 20.75 \\
         &  & $\Delta B_{\phi}$ & 178099 & -0.15 & 19.45 & 0.08 & 23.52 \\
         &  & $\Delta F$ & 178099 & 0.37 & 6.67 & 0.14 & 7.36 \\
        \cmidrule{1-8}
        \multirow[c]{8}{*}{Swarm-A} & \multirow[c]{4}{*}{Non-polar} & $\Delta B_r$ & 190878 & 0.09 & 2.78 & 0.04 & 3.96 \\
         &  & $\Delta B_{\theta}$ & 190878 & -0.05 & 3.53 & -0.06 & 3.89 \\
         &  & $\Delta B_{\phi}$ & 190878 & 0.03 & 5.32 & 0.06 & 5.84 \\
         &  & $\Delta F$ & 190878 & -0.09 & 2.62 & -0.26 & 3.56 \\
        \cmidrule{2-8}
         & \multirow[c]{4}{*}{Polar} & $\Delta B_r$ & 118640 & -0.04 & 5.58 & -0.08 & 7.42 \\
         &  & $\Delta B_{\theta}$ & 118640 & 0.11 & 15.89 & 0.11 & 19.65 \\
         &  & $\Delta B_{\phi}$ & 118640 & 0.05 & 18.83 & 0.39 & 23.28 \\
         &  & $\Delta F$ & 118640 & -0.22 & 5.18 & -0.75 & 7.27 \\
        \cmidrule{1-8}
        \multirow[c]{8}{*}{Swarm-B} & \multirow[c]{4}{*}{Non-polar} & $\Delta B_r$ & 192185 & -0.02 & 2.71 & -0.08 & 3.89 \\
         &  & $\Delta B_{\theta}$ & 192185 & -0.12 & 3.46 & -0.10 & 3.79 \\
         &  & $\Delta B_{\phi}$ & 192185 & -0.01 & 5.27 & 0.04 & 5.78 \\
         &  & $\Delta F$ & 192185 & -0.00 & 2.52 & -0.21 & 3.44 \\
        \cmidrule{2-8}
         & \multirow[c]{4}{*}{Polar} & $\Delta B_r$ & 119224 & -0.12 & 5.18 & -0.13 & 6.83 \\
         &  & $\Delta B_{\theta}$ & 119224 & 0.29 & 15.85 & 0.23 & 19.54 \\
         &  & $\Delta B_{\phi}$ & 119224 & 0.11 & 18.61 & 0.44 & 22.99 \\
         &  & $\Delta F$ & 119224 & 0.05 & 4.70 & -0.48 & 6.62 \\
        \cmidrule{1-8}
        \multirow[c]{8}{*}{Swarm-C} & \multirow[c]{4}{*}{Non-polar} & $\Delta B_r$ & 194920 & 0.06 & 2.79 & -0.00 & 3.93 \\
         &  & $\Delta B_{\theta}$ & 194920 & -0.11 & 3.54 & -0.13 & 3.90 \\
         &  & $\Delta B_{\phi}$ & 194920 & 0.03 & 5.29 & 0.07 & 5.82 \\
         &  & $\Delta F$ & 194920 & -0.01 & 2.61 & -0.17 & 3.54 \\
        \cmidrule{2-8}
         & \multirow[c]{4}{*}{Polar} & $\Delta B_r$ & 121355 & 0.01 & 5.54 & -0.02 & 7.38 \\
         &  & $\Delta B_{\theta}$ & 121355 & 0.10 & 15.82 & 0.09 & 19.62 \\
         &  & $\Delta B_{\phi}$ & 121355 & 0.05 & 18.77 & 0.41 & 23.19 \\
         &  & $\Delta F$ & 121355 & -0.05 & 5.15 & -0.58 & 7.24 \\
        \bottomrule
    \end{tabular}
\end{table}
}%
The RMS values of \modela{} are generally lower than those of \modelref{}, especially in the case of horizontal and scalar residuals at polar latitudes. For example, the RMS values of $\Delta B_\theta$, $\Delta B_\phi$, and $\Delta F$ for CHAMP in the polar regions are \SI{17.35}{\nT}, \SI{18.89}{\nT}, and \SI{6.20}{\nT} for \modela{} and \SI{21.07}{\nT}, \SI{23.14}{\nT}, and \SI{8.52}{\nT} for the reference model, respectively, which corresponds to a reduction of approximately \SI{20}{\percent}. This suggests that co-estimating an ionospheric field improves the fit to the data, in particular, at high latitudes. To further characterise the improvement in the data fit, we investigated median scalar residuals as a function of QD latitude and MLT, which we computed in cells of approximately equal area using a HEALPix \parencite[]{Gorski2005} grid in QD/MLT coordinates over the entire globe. In Fig.~\ref{fig:scalar_residual_maps} we compare these maps of median scalar residuals for \modela{} and the reference model.
Looking at the polar views for the reference model, median scalar residuals are organised into a pattern of two crescent-shaped cells of positive and negative values around each magnetic pole, which reflects the well-known two-cell current system in the polar ionosphere \parencite[e.g.][]{Dungey1961}. Likewise, in the global view for the reference model, the strongly negative values of the median scalar residuals on the dayside at the dip-equator and slightly less negative values at mid-latitudes are associated with the solar quiet current system and the equatorial electrojet \parencite[e.g.][]{Yamazaki2016}. In \modela{}, the median value of the scalar residuals in QD/MLT coordinates are dramatically reduced not only at polar latitudes, where the AMPS approach is expected to work best, but also at low and mid-latitudes, in particular, close to the dip equator on the dayside. The remaining patterns are relatively weak and could possibly be captured by using a higher truncation degree of the ionospheric field model. The fact that the patterns found for the reference model are largely absent for \modela{}, in the bottom panel of Fig.~\ref{fig:scalar_residual_maps}, shows that our approach accounts for previously unmodelled signals associated with ionospheric current systems in the residuals.

\begin{figure*}
    \centering
    \includegraphics{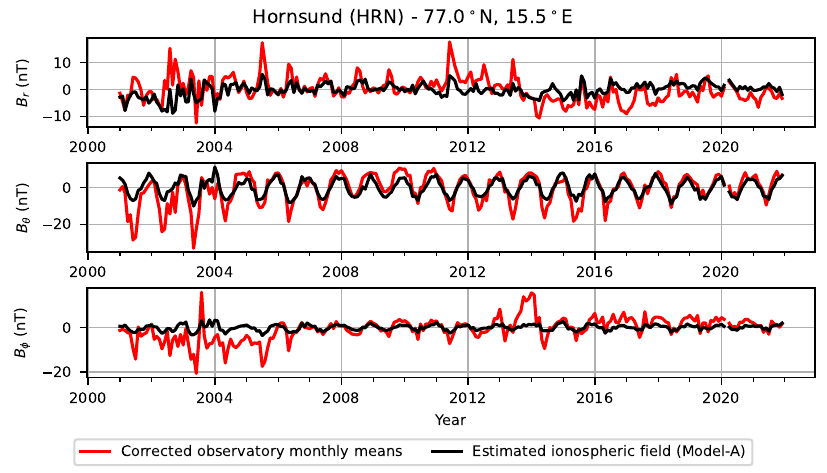}
    \caption{Monthly mean values of the quiet-time ionospheric field extracted from the records of the magnetic observatory in Hornsund, Svalbard in Norway (red) and those given by \modela{} at the same location (black). Note that we ignored the effect of induction during the downward continuation of our ionospheric field model.}
    \label{fig:HRN_timeseries}
\end{figure*}

Since the models in this study are derived only from satellite magnetic data, we can test how well time variations are modelled in comparison to independent observations made by ground-based magnetic observatories of the International Real‐time Magnetic Observatory Network (INTERMAGNET). These data are available as hourly mean values at the World Data Centres for Geomagnetism in Edinburgh and have been quality checked as explained in \cite{Macmillan2013}. In Fig.~\ref{fig:HRN_timeseries}, for example, we compare monthly means of the ionospheric field given by \modela{} compared with those recorded at the ground-based magnetic observatory in Hornsund (HRN), located at \SI{77.0}{\degree}N, \SI{15.5}{\degree}E on Svalbard (Norway) near the northern edge of the auroral oval.
For this comparison, we extracted monthly means of the ionospheric field from the timeseries of hourly means at HRN by applying the quiet-time selection in Sect.~\ref{sec:magnetic_observations_and_data_selection}, removing the internal and magnetospheric field estimates given by \modela{}, and centring the corrected timeseries with the component-wise average to remove the remaining crustal field. Concerning the modelled ionospheric field, we produced monthly averages using the hourly estimates of the poloidal ionospheric field of \modela{} at the times of the quiet-time HRN timeseries after downward continuing this part of the model below the reference height to the Earth's surface. For this, we assumed that the poloidal ionospheric field below the reference height can be represented through an external potential whose radial field component is continuous across the reference height. Note that the toroidal ionospheric part of the model does not contribute to the ionospheric field on ground since it does not exist inside the non-conducting part of the atmosphere. Fig.~\ref{fig:HRN_timeseries} shows that \modela{} closely follows yearly and slower variations of the HRN timeseries, especially the fit to the southward component is encouraging. However, \modela{} certainly underestimates the peak values seen in the observatory timeseries, for example, for those in 2003, and cannot reproduce the more dynamic time periods between 2001 and 2005, and between 2012 and 2016, which are associated with solar maximum conditions. This shows that our approach sensibly models the typical variations of the ionospheric field but is not able to reproduce the dynamic ionospheric field produced by local currents above the observatory.

\begin{figure*}
    \centering
    \includegraphics{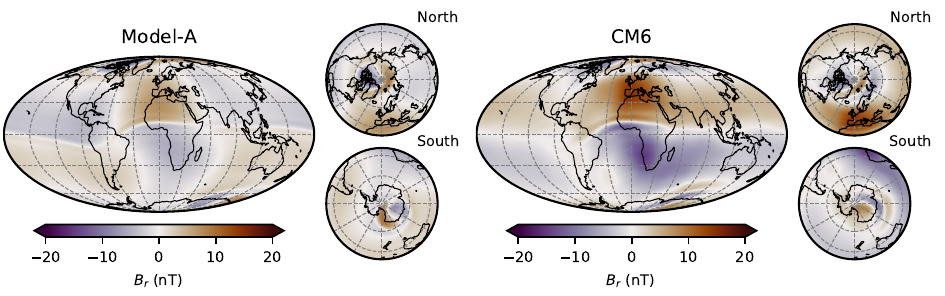}
    \caption{Radial component of the ionospheric magnetic field from \modela{} (left) and CM6 (right) at \SI{450}{km} altitude during noon in Greenwich on March 21, 2018. Note that we show the sum of inducing and induced parts of the ionospheric field from CM6.}
    \label{fig:ionospheric_field}
\end{figure*}

To document our estimated ionospheric field at mid and low latitudes, we show in Fig.~\ref{fig:ionospheric_field} the radial component of the ionospheric magnetic field from \modela{} and from CM6 \parencite[]{Sabaka2020}, the sum of primary and secondary parts, at satellite altitude at \SI{450}{\kilo\meter} during noon in Greenwich on March 21, 2018.
The overall pattern of the ionospheric field from \modela{} at low and mid latitudes is similar to the CM6 model. The radial field is stronger around local noon north and south of the magnetic dip equator and the geometry broadly follows that of the internal field. However, there are important differences between \modela{} and CM6. At mid and low latitudes the amplitude of the radial field from \modela{} is much weaker than for CM6 on the dayside, while it is comparable on the nightside but with different signs in the northern and southern hemispheres. At high latitudes the two cell pattern around the magnetic poles is clearly visible for \modela{}, while the pattern is weaker and smeared out for CM6, in particular in the southern hemisphere. This shows that our ionospheric field model captures the basic Solar-quiet pattern but has limitations on the dayside at non-polar latitudes. Note that the ionospheric field from \modela{} becomes more similar to CM6 at low and mid latitudes by relaxing the regularisation imposed on the divergence-free sheet currents associated with this part of the model.

\subsection{Ionospheric currents during geomagnetic quiet-time conditions}

In this section we report the spatial structure of the estimated currents from \modela{} and their response to changes in the external driving.

\begin{figure*}
    \centering
    \includegraphics{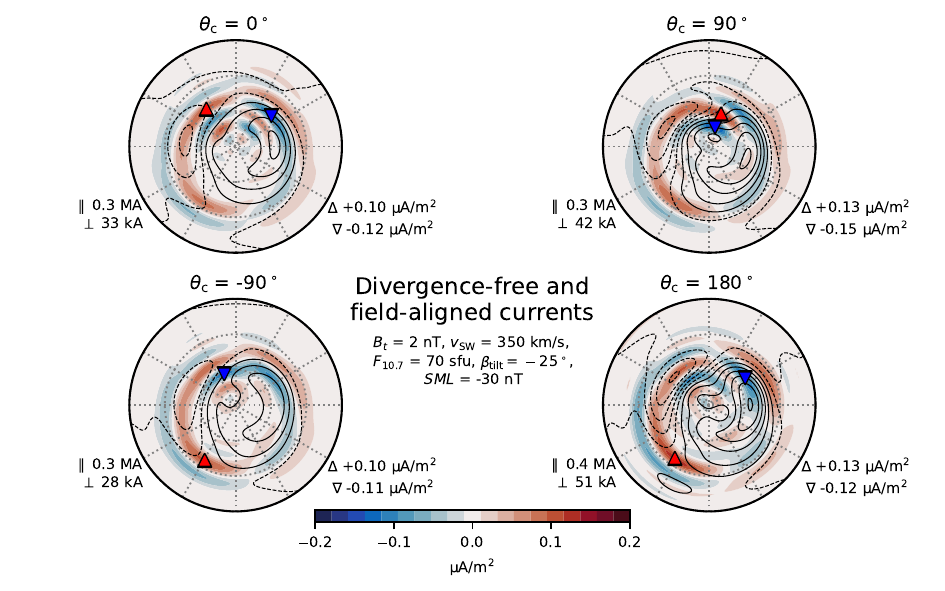}
    \caption{Divergence-free and field-aligned current densities in the north polar region for different clock angles. Each panel shows the current densities in QD/MLT coordinates above \SI{60}{\degree} QD latitude with noon at the top, dawn on the right, midnight at the bottom and dusk on the left. The contours show the potential of the divergence-free sheet current density in steps of \SI{5}{\kilo\ampere} (solid for positive and dashed for negative values) and the colours indicate the field-aligned current density. The location of the largest ($\bigtriangleup$) and smallest ($\bigtriangledown$) field-aligned current densities are indicated with the coloured triangles, and their strength is given in the lower right corner of each panel. The total field-aligned current ($\parallel$) and the divergence-free part of the sheet current ($\perp$) flowing between the maximum and minimum of $\psi^\mathrm{df}$, poleward of \SI{60}{\degree} QD latitude, are given in the lower left corner. The dotted lines indicate QD parallels at \SI{10}{\degree} and MLT meridians at \SI{2}{\hour} intervals. This figure is similar in form to those originally presented by \cite{Laundal2018} for the AMPS model.}
    \label{fig:climatology}
\end{figure*}

In Fig.~\ref{fig:climatology} we show polar views of the divergence-free and field-aligned current densities in QD/MLT coordinates as a function of clock angle for the quiet solar wind conditions represented in the dataset during winter in the northern hemisphere.
For all clock angles, the divergence-free sheet currents circulate in two cells, roughly separated by the noon-midnight meridian, whereas the field-aligned currents form concentric patterns with the centre slightly offset from the magnetic pole towards midnight, known as R1 and R2 currents \parencite[]{Iijima1976,Iijima1978}. For northward IMF ($\theta_\mathrm{c}=\SI{0}{\degree}$) the cell of the divergence-free sheet currents in the dawn sector is slightly more pronounced than the one in the dusk sector, and the maxima of the field-aligned currents are located close to noon. When the IMF rotates southward, the currents generally gain in strength. However, a clockwise rotation to $\theta_\mathrm{c}=\SI{90}{\degree}$ leads to stronger currents than a counter-clockwise rotation to $\theta_\mathrm{c}=\SI{-90}{\degree}$, which corresponds to an asymmetry in the currents that depends on the y-component of the IMF. The currents are strongest when the IMF is southward ($\theta_\mathrm{c}=\SI{180}{\degree}$) with a maximum of downward field-aligned currents in the noon-dawn sector and a maximum of upward field-aligned currents in the midnight-dusk sector. The case of southward oriented IMF, however, should be interpreted with caution since it is poorly represented in the data due to the chosen data selection (see panel for the clock angle in the top centre of Fig.~\ref{fig:external_parameter_distribution}).

\begin{figure*}
    \centering
    \includegraphics{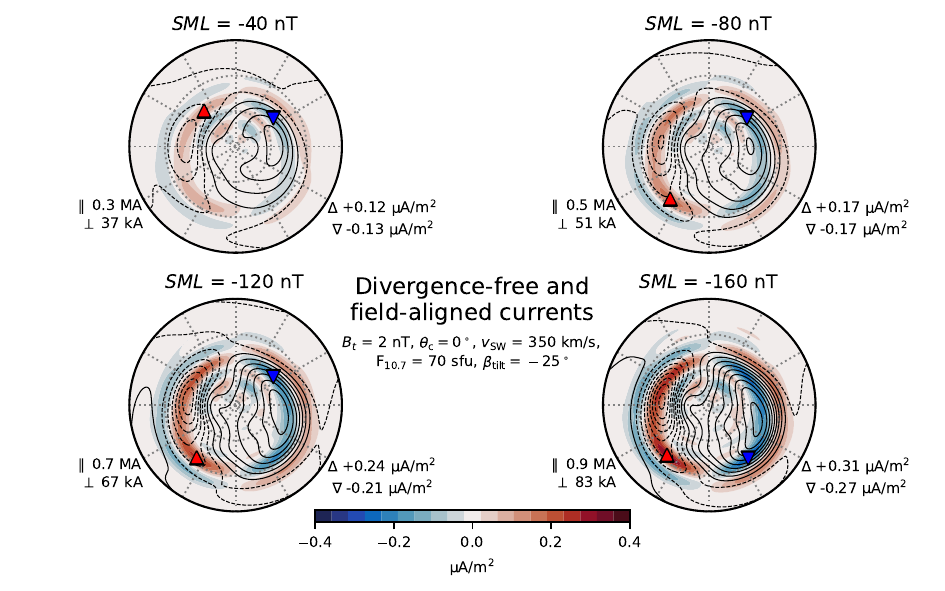}
    \caption{Divergence-free and field-aligned current densities in the north polar region for different values of the $\mathit{SML}$ index and purely northward IMF. The figure is otherwise identical to Fig.~\ref{fig:climatology}.}
    \label{fig:climatology_sml}
\end{figure*}

Fig.~\ref{fig:climatology_sml} shows the divergence-free and field-aligned current densities for the same conditions as in Fig.~\ref{fig:climatology} but in dependence of the $\mathit{SML}$ index and only for purely northward IMF.
With decreasing $\mathit{SML}$ index, corresponding to an increase in auroral activity, the divergence-free sheet currents and the field-aligned currents grow in strength. The locations where the field-aligned currents are strongest move from near noon ($\mathit{SML}=\SI{-40}{\nT}$) to the midnight sector ($\mathit{SML}=\SI{-160}{\nT}$), reaching a configuration in which an upward directed current dominates the pre-midnight and a downward current the post-midnight sectors. This pair of strong downward and upward field-aligned currents centred on midnight agrees well with the current wedge that is thought to exist during substorms \parencite[]{McPherron1973,Kepko2015}. Again, however, the case for $\mathit{SML}=\SI{-160}{\nT}$ should be considered an extrapolation given that there was a relatively small amount of data with large negative values of the $\mathit{SML}$ index available during the modelling thanks to the quiet-time data selection (for the distribution of the $\mathit{SML}$ see lower right panel in Fig.~\ref{fig:external_parameter_distribution}).

For interpreting the variations of the patterns in Figs.~\ref{fig:climatology} and \ref{fig:climatology_sml} in terms of physical processes, it is worth mentioning that the terms in the ionospheric magnetic field parametrization related to the $\mathit{SML}$ index and the $\epsilon$ coupling function probably compete to some extent for the same signal. So a part of the structure which may go into the $\epsilon$ terms, if the $\mathit{SML}$ index was not included in the parametrization, may now be contained in the $\mathit{SML}$ terms. For example in Fig.~\ref{fig:climatology}, the changes due to variations in the clock angle may be less than in the AMPS model, which did not include the $\mathit{SML}$ index, because some of the signal for southward IMF conditions is now contained in the $\mathit{SML}$ terms.

\begin{figure*}
    \centering
    \includegraphics{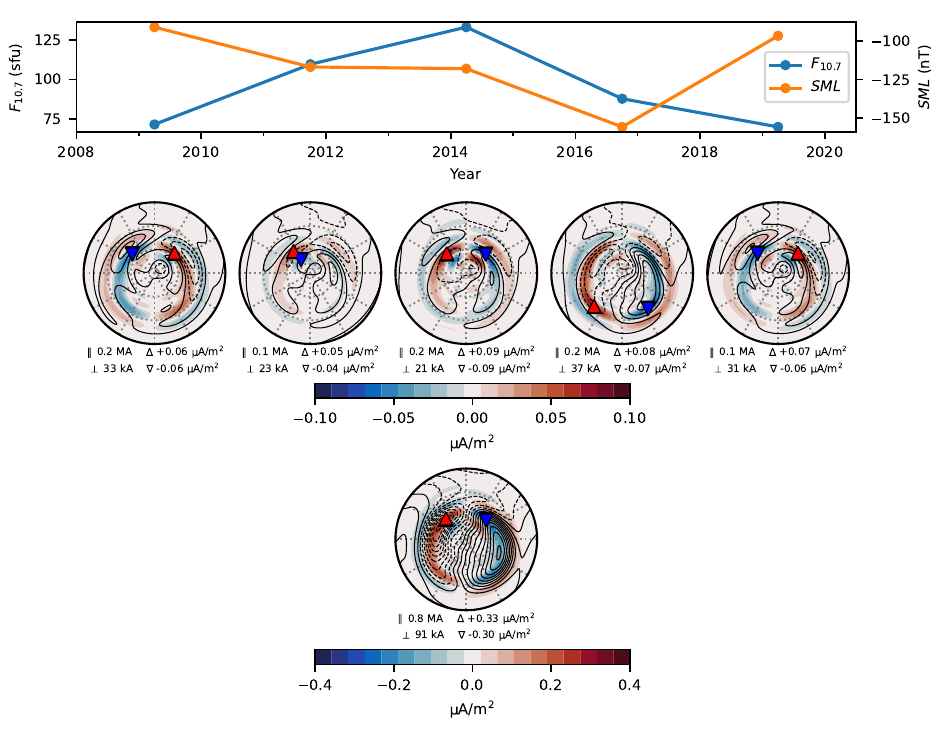}
    \caption{Average $F_{10.7}$ and $\mathit{SML}$ indices (top row), and divergence-free and field-aligned current densities (middle row) in the north polar region successively averaged over 2.5 year periods after removal of the average current density (bottom row) over the solar cycle from 2008 to 2020.5. The markers on the curves of the indices are placed at the midpoints of the 2.5 year intervals used for the averaging, and the contours of the potential of the divergence-free sheet current density are shown in steps of \SI{5}{\kilo\ampere}.}
    \label{fig:currents_solar_cycle_average}
\end{figure*}

Finally, to illustrate the ability of the climatological approach to allow solar cycle changes in the ionospheric currents, we produced a sequence of plots, similar to Figs.~\ref{fig:climatology} and \ref{fig:climatology_sml}, that show the current densities in the northern polar region averaged in time over successive 2.5 year intervals between 2008.0 and 2020.5. Specifically, we averaged the currents over the intervals 2008.0--2010.5 (solar minimum), 2010.5--2013.0 (ascending solar cycle), 2013.0--2015.5 (solar maximum), 2015.5--2018.0 (descending solar cycle), and 2018.0--2020.5 (again solar minimum) to illustrate different phases of the solar cycle. To further emphasise the changes in the currents, we removed the average current density for the entire solar cycle shown here. Fig.~\ref{fig:currents_solar_cycle_average} shows this sequence of plots and the removed average current density along with the $F_{10.7}$ and $\mathit{SML}$ indices, averaged using the same 2.5 year intervals.
The $F_{10.7}$ index reaches a maximum around 2014, indicating solar maximum, whereas the $\mathit{SML}$ index has a minimum later, close to 2017, coinciding with the descending phase of the solar cycle. Turning to the field-aligned currents, differences with respect to the solar cycle average mostly occur in the noon sector with the exception of the descending phase, when the differences are most prominent in the midnight sector. The differences in the divergence-free sheet currents with respect to the solar cycle average are generally more complex. However, noteworthy is the two-cell pattern that is visible during the descending phase of the solar cycle, which is consistent with the enhancement of this pattern for decreasing values of the $\mathit{SML}$ index as shown in Fig.~\ref{fig:climatology_sml}. Overall, we conclude that our approach is able to capture at least part of the changes that occur in the strength and appearance of polar ionospheric currents during the solar cycle.

\subsection{Core field secular variation at polar latitudes}

We demonstrated above that the co-estimation of the polar ionospheric currents helps with accounting for previously unmodelled signals in the residuals. We now turn to the impact of co-estimating the ionospheric field on the estimated core field, by considering differences between the internal field estimates of \modela{} and the reference model.

\begin{figure*}
    \centering
    \includegraphics{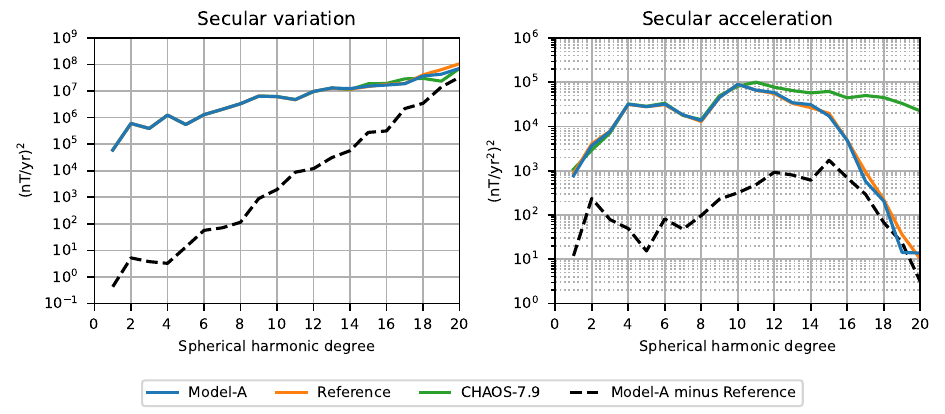}
    \caption{Spatial power spectrum of the SV (left) and SA (right) at the CMB in 2019.0 from \modela{} (blue), \modelref{} (orange), CHAOS-7.9 (green), and the difference between \modela{} and \modelref{} (black dashed).}
    \label{fig:power_spectrum}
\end{figure*}

\begin{figure*}
    \centering
    \includegraphics{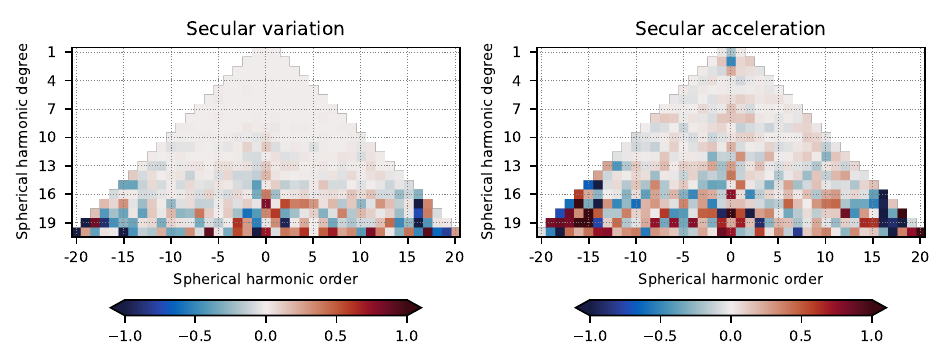}
    \caption{Sensitivity matrix of the SV (left) and the SA (right) in 2019.0 from \modela{} with respect to \modelref{}.}
    \label{fig:sensitivity}
\end{figure*}

In Fig.~\ref{fig:power_spectrum} we show the spatial power spectra of the Secular Variation (SV) and Secular Acceleration (SA) at the CMB in 2019.0 from \modela{}, \modelref{}, CHAOS-7.9 , and the difference between \modela{} and \modelref{}.
The SV spectra are very similar at low spherical harmonic degree, causing the curves for the models to overlap in the plot. Above degree 13 the spectra deviate more clearly but continue to stay closely together as the degree increases. The SA spectra increase with spherical harmonic degree and reach a maximum at degree 10. Above degree 10 the spectra decrease but much steeper for \modela{} and \modelref{} than for CHAOS-7.9. This difference in the high-degree SA is the result of the temporal regularisation used in CHAOS-7.9, which is tapered to allow for more power at high degrees. However, the taper was not applied in the models of this study. Since the spectra from \modela{} and \modelref{} are very similar, we show in Fig.~\ref{fig:sensitivity} the sensitivity matrix of the SV and the SA in 2019.0 from \modela{} with respect to \modelref{}.
The sensitivity matrix is defined as the coefficient-wise difference between recovered spherical harmonic coefficients and chosen target coefficients, here the coefficients from \modelref{}, normalised with the mean amplitude of the target coefficients at degree $n$ \parencite[e.g.][]{Sabaka2013}. The SV coefficients of \modela{} and \modelref{} are very similar at low spherical harmonic degree. However, the sensitivity increases above degree 14 in particular for near-zonal ($m \approx 0$) and near-sectorial ($m \approx n$) coefficients. A similar pattern can be observed in the sensitivity matrix of the SA, but the numerical values are overall larger. Noteworthy are relatively strong sensitivity values for the zonal SA coefficients at degree 2 and 3.

\begin{figure*}
    \centering
    \includegraphics{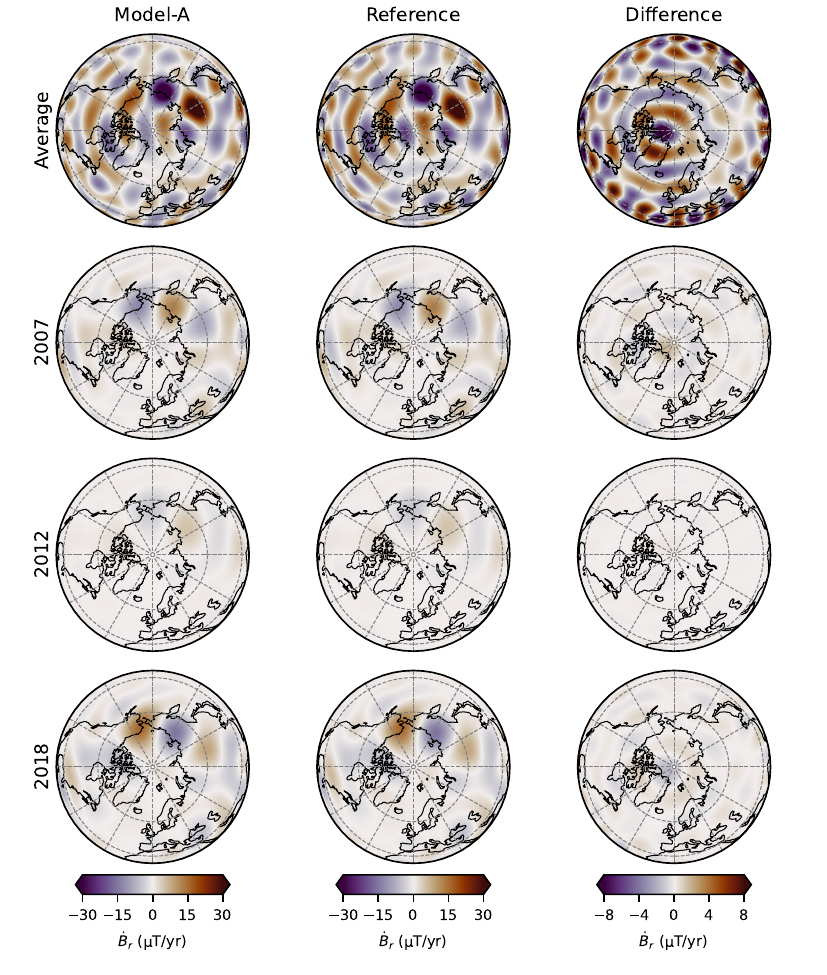}
    \caption{Snapshots of the radial SV for $n \leq 19$ in the north polar region at the CMB as given by \modela{} (left column), the reference model (middle column) and their difference (right column) in 2007.0 (second row), 2012.0 (third row), 2018.0 (fourth row), after removing the snapshot average (first row). The projection is orthographic. The dashed lines show geographic parallels and meridians at \SI{30}{\degree} intervals. Note the change in colour scale for the difference plots.}
    \label{fig:sv_panels}
\end{figure*}

The spatial power spectra and the sensitivity matrix for the SV show that appreciable differences between \modela{} and the reference model can be expected at high spherical harmonic degrees. To examine this we plot in Fig.~\ref{fig:sv_panels} snapshots of the radial SV for \modela{} and \modelref{} up to degree 19 at the CMB in the north polar region in 2007.0, 2012.0 and 2018.0, after removing the snapshot average to emphasise changes in time.
The average radial SV for \modela{} shows small-scale flux patches of positive and negative polarity, with a distinct pair of patches located over Siberia. This pair is also visible in the average map for the reference model but is less prominent due to relatively strong patches west of Greenland, which are slightly elongated in longitude. For \modela{}, the flux patches around that area are less extended in longitude, showing that the co-estimation of the ionospheric field leads to better focused patches of SV at the CMB. This can also be seen in the difference between the average radial SV from \modela{} and \modelref{}, which exhibits near-zonal features of radial SV around the north pole. Despite this improvement in \modela{}, the presence of a weak pattern of stripes in the SV along geographic parallels over North America may indicate that the SV at high latitudes is still contaminated by the ionospheric field at high degree.

The snapshots showing the deviation of the radial SV from the average reveal that the three non-axisymmetric SV flux patches over Siberia and Alaska intensify for \modela{} over the model time interval, and similarly for \modelref{}, by an almost linear trend. The lack of structure in the difference between \modela{} and \modelref{} for each of these snapshots shows that the difference between the radial SV from \modela{} and from \modelref{} is mostly static.

Overall, it seems that the co-estimation of the ionospheric magnetic field leads to less contaminated models of the SV in the north polar region that are derived from magnetic vector data at all latitudes and local times. It is clear that the distinctive non-axisymmetric radial SV flux patches of alternating sign over Siberia and Alaska, found in earlier geomagnetic field models and used for inferring accelerating jets of core flow \parencite[]{Livermore2017}, persist even when polar ionospheric currents are estimated. A clear limitation in examining the high degree CMB SV in these models is that they are strongly smoothed in time. The effect of reducing the strong temporal regularisation will be investigated in the next section.

\subsection{Relaxing the temporal regularisation of the internal field model}

Our approach for modelling the geomagnetic field relies on regularisation to smooth the spatio-temporal complexity of the model and, thus, ensures the convergence of the estimation procedure. The regularisation also allows control over magnetic signals in the data that are not accurately parametrized. However, in the case of the time-dependent internal field model, the temporal regularisation severely degrades the resolution in time, in particular, of the high-degree spherical harmonics, which limits studies of the dynamics of the Earth's interior. In this section, we therefore explore a model in which, building on \modela{}, we in addition considerably relaxed the temporal regularisation of the internal field model. More specifically, we reduced the temporal regularisation parameters $\lambda_t$, $\lambda_{t_s}$ and $\lambda_{t_e}$ by a factor of \num{80} to \num{0.0125}, \num{1.25e-4} and \num{1.25e-4}, respectively, to produce a weakly-regularised version of \modela{}, called \modelb{}.

\begin{figure*}
    \centering
    \includegraphics{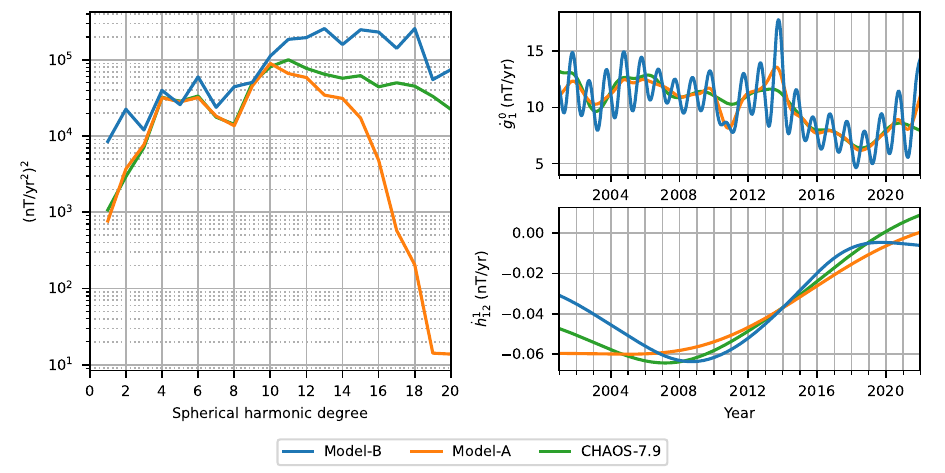}
    \caption{Spatial power spectrum of the SA at the CMB in 2019.0 (left) and timeseries of the SV coefficients $\dot{g}_1^0$ (top right) and $\dot{h}_{12}^1$ (bottom right) for \modelb{} (blue), \modela{} (orange) and CHAOS-7.9 (green).}
    \label{fig:spectrum_gauss_timeseries}
\end{figure*}

On the left of Fig.~\ref{fig:spectrum_gauss_timeseries}, we show the spatial power spectrum of the SA at the CMB in 2019.0 for \modelb{}, \modela{} and CHAOS-7.9.
The spectra show that there is significantly more power in the SA of \modelb{} at all spherical harmonic degrees. Most striking is the increase in SA power for \modelb{} above degree 10,  while the power for \modela{} sharply decreases. But also noteworthy is the enhanced power below degree 4 for \modelb{}. For comparison we also show the spectrum for CHAOS-7.9, which overlaps below degree 10 with that for \modela{} but decreases more slowly as the degree further increases. On the right of Fig.~\ref{fig:spectrum_gauss_timeseries}, we show example timeseries of SV coefficients for the three models. The timeseries of $\dot{g}_1^0$ reveals a distinct annual oscillation for \modelb{}. Similarly, we found annual oscillations in the timeseries of zonal and low-order coefficients for other low-degree spherical harmonics. They are investigated further below. We emphasise that these annual oscillations are not an effect of the applied data selection, which does not vary with season. The oscillations are not as apparent in the coefficient timeseries of \modela{} and CHAOS-7.9, due to the relatively strong temporal regularisation. Compared to $\dot{g}_1^0$, the $\dot{h}_{12}^1$ timeseries is much smoother for all three models and there is no oscillatory behaviour with a period close to one year visible. So \modelb{} is still quite strongly smoothed in time at high degree. This is by construction due to the chosen regularisation norm. Note that we chose to show the SA spectrum in 2019.0 in Fig.~\ref{fig:spectrum_gauss_timeseries} because the rate of change of the SV maximises around that time for \modelb{}. In fact, we find that the spectrum for \modelb{} varies significantly with time below degree 8, reflecting the annual variations found in the low-order and low-degree SV coefficients. These temporal variations are well known and have been handled by other modelling efforts in different ways, by using regularisation [e.g., CHAOS models \parencite[]{Finlay2020}, GRIMM \parencite[]{Lesur2008}, or CM \parencite[]{Sabaka2020}], low resolution basis functions [e.g., POMME \parencite[]{Maus2006}, CovObs \parencite[]{Huder2020}], or by including other internal sources [e.g., Kalmag models \parencite[]{Baerenzung2022}, or sequential models of \cite{Ropp2020}].

To better characterise the annual oscillations found in the low spherical harmonic coefficients of the time-dependent internal field model of \modelb{}, we performed a principal component analysis (PCA) of the difference between \modelb{} and \modela{}. The PCA is a data-based tool for extracting spatio-temporal patterns and has previously been applied to the magnetic data from ground-based observatories \parencite[e.g.][]{Shore2016} and magnetic observations made by satellites \parencite[e.g.][]{Domingos2019,Saturnino2021}. For the analysis, we generated timeseries of vector components of the internal time-dependent field from each model at the centre points of equal-area pixels \parencite[]{Gorski2005}, covering the entire Earth's surface at approximately \SI{2}{\degree} resolution (10800 pixels), in spherical geocentric coordinates using a sampling rate of one sample per month (201 samples in time). Here, we omitted the period during the CryoSat-2 data, from the middle of 2010 to the end of 2013, when \modelb{} varies more strongly in time than during CHAMP and \swarm{}, and we omitted the first and last 6 months of the model time interval, when \modelb{} shows sharp time variations due to the weaker data constraint at the model endpoints. At each pixel, by subtracting the vector timeseries of \modela{} from the one of \modelb{}, we obtained timeseries of component-wise differences, which we arranged as columns in a matrix $\mathbf{X}$ of size $201 \times 32400$. Finally, we centred $\mathbf{X}$ by removing the column-wise mean value, $\bar{\mathbf{X}}$, such that
\begin{equation}
    \tilde{\mathbf{X}} = \mathbf{X} - \bar{\mathbf{X}}.
\end{equation}

For a data matrix such as $\tilde{\mathbf{X}}$, PCA can be used to find a finite set of modes that maximise the variance of the data in time and are mutually orthogonal. The $i$th mode obtained through the PCA consists of the Empirical Orthogonal Function (EOF), $\mathbf{v}_i$, which represents the spatial pattern of the time variation, and the principal component (PC) $\mathbf{y}_i = \tilde{\mathbf{X}}\mathbf{v}_i$, which is the timeseries of variance $\sigma^2_i$. These modes are typically sorted in decreasing order of variance and can be used to reconstruct the data matrix through
\begin{equation}
    \tilde{\mathbf{X}} = \sum_i \mathbf{y}_i\mathbf{v}^\mathrm{T}_i,
\end{equation}
or, if a subset of modes is chosen, to perform a partial reconstruction.

\begin{figure*}
    \centering
    \includegraphics{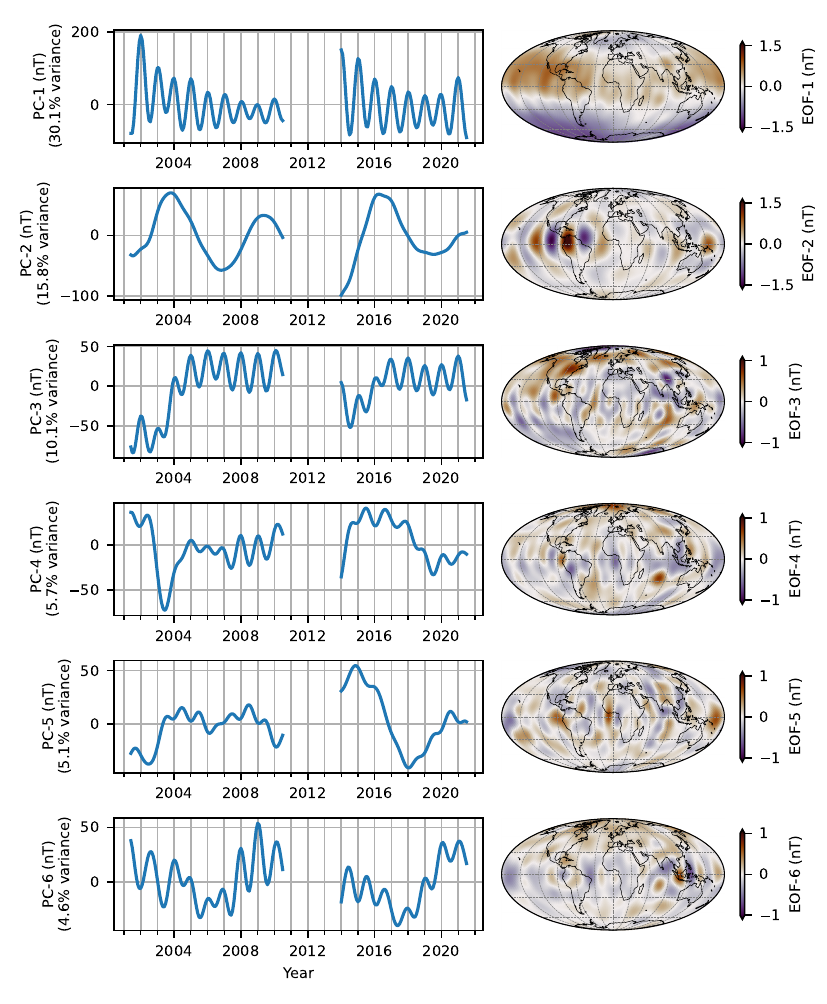}
    \caption{PCs (left) and radial part of the corresponding EOFs (right) for the first six modes found by the PCA. The EOFs are scaled with the square-root of the mode variance to indicate the relative importance.}
    \label{fig:pca}
\end{figure*}

We applied the PCA on $\tilde{\mathbf{X}}$ and obtained 201 modes. But only a small number of modes are needed to explain most of the variance in the data. In Fig.~\ref{fig:pca}, we show the PCs and the radial part of the EOFs for the first six modes, which account for \SI{70}{\percent} of the variance (13 modes account for \SI{90}{\percent}).
The first PC, PC-1, is a modulated annual oscillation for which the amplitude maximises around 2002 and 2014, and minimises around 2008 and 2019. The corresponding EOF, EOF-1, consists of a large-scale pattern that varies mostly in latitude, similar in appearance to the spherical harmonic function $Y_2^0$ except that the zero lines in latitude are shifted slightly northward. Given the spatio-temporal behaviour, we can assume that the first mode is responsible for most of the annual oscillations in the coefficient timeseries, as shown for example by $\dot{g}_1^0$ in Fig.~\ref{fig:spectrum_gauss_timeseries} (phase shift between $\dot{g}_1^0$ and PC-1 reflects the difference in time derivative between the two). The amplitude modulation of PC-1 suggests a dependency on the solar cycle since the maxima in the amplitude roughly coincide with the times of solar maximum. PC-2 varies more slowly compared to PC-1 and peaks approximately every 3 years. EOF-2 shows patches of opposite sign that are centred on the geographic equator around Central America and the western Pacific Ocean. The location and appearance of these patches could indicate that the second mode is related to changes in the core field since geomagnetic impulses have been reported around these areas \parencite[]{Olsen2007,Chulliat2014,Torta2015,Finlay2020}. The third PC, PC-3, combines a slow variation that peaks around 2002 and 2014 with an annual oscillation, which is much weaker in amplitude compared to PC-1 but likely also of external origin. EOF-3 exhibits a large number of small-scale positive and negative patches. The remaining PCs (PC-4 through PC-6) show an oscillatory behaviour with a period close to one year and the corresponding EOFs exhibit a wide range of patterns, which are difficult to interpret.

Based on the results of the PCA, we tried to implement a post-processing step that removes the modes that we assume are the result of a leakage of the external field into the internal field model as is the case, for example, for the first mode, since it is a global annual oscillation with an apparent solar cycle dependency. Successful removal of these modes would provide a smoother timeseries of the coefficients of the time-dependent internal field model, which we could use to analyse the SA. Unfortunately, this approach did not work in practice since the PCA is not able to isolate the annual oscillations in terms of a single mode. Instead, we found that these oscillations are visible in most, if not all, of the modes and cannot be removed to a satisfactory level, including by further relaxing the temporal regularisation, which makes the annual oscillations clearer. Nonetheless, we find the PCA provides clear insight into the signals entering internal field models as the temporal regularisation is relaxed.

\begin{figure*}
    \centering
    \includegraphics{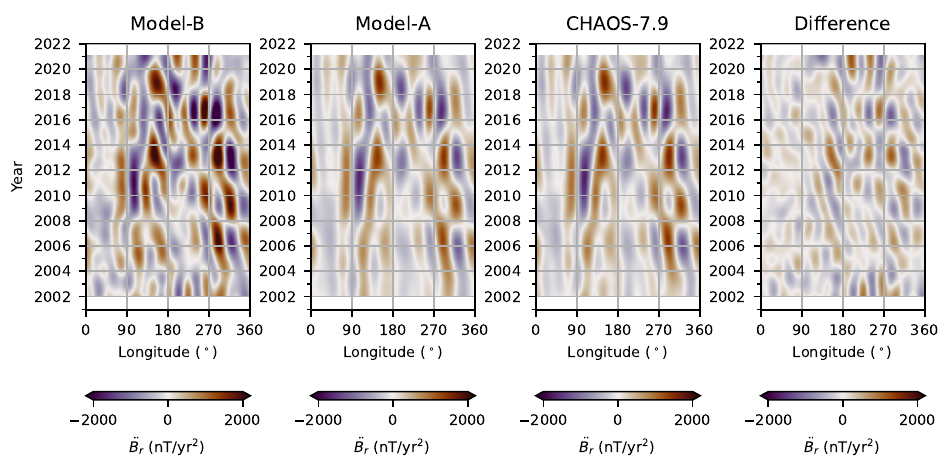}
    \caption{Time-longitude plot of the radial SA up to degree 10 along the geographic equator at the CMB as given by repeated computation of annual differences of the radial field from \modelb{}, \modela{}, CHAOS-7.9, and the radial field difference between \modelb{} and CHAOS-7.9.}
    \label{fig:sa_time_longitude}
\end{figure*}

As a simpler alternative to removing PCA modes, we resorted to filtering out the annual oscillations by computing centred annual differences of monthly values of the internal field at the CMB as given by \modelb{} to find the SV and, by repetition, the SA. Fig.~\ref{fig:sa_time_longitude} shows time-longitude plots of the obtained radial SA up to degree 10 for \modelb{} on the left and, by the same computation, for \modela{} and CHAOS-7.9 in the middle, and the difference between \modelb{} and CHAOS-7.9 on the right.
We see that, in comparison to \modela{} and also CHAOS-7.9, which is similar to \modela{} in Fig.~\ref{fig:sa_time_longitude}, there is more power in the SA of \modelb{}. The patterns have sharper edges and there is generally more structure, which indicates an improved temporal resolution. This is especially apparent during the CHAMP period, until 2010, and in the longitude interval between \SI{0}{\degree} and \SI{90}{\degree} for the entire model time span. We produced a similar plot for a weakly regularised version of \modelref{} and found that it looks very similar to Fig.~\ref{fig:sa_time_longitude}. We interpret this to mean that the increase in resolution is mostly due to the reduced temporal regularisation and not due to the co-estimation of the ionospheric field model. We acknowledge that computing annual differences has the caveat of removing genuine internal field signals that have frequencies which are integer multiples of one oscillation per year. Further work is clearly needed on better methods to remove the annual signal.

To summarise, we find that reducing the temporal regularisation of the internal field model causes signals which we suspect are of external origin to leak into the estimated internal field despite the co-estimation of an AMPS-type model of the ionospheric field. It is possible that our approach to parametrize the ionospheric field lacks terms that can account for the signals like those seen in the first and third mode identified by the PCA (see discussion in Sect.~\ref{sec:discussion}). To ensure that the co-estimation of the ionospheric field model is not, in fact, introducing artefacts into the internal field model, we performed the same analysis of principal components but using the reference model. We found that the first two modes are similar, which suggests that these modes originate from genuine signals in the magnetic dataset used in this study.

\section{Discussion}
\label{sec:discussion}

Our results show that the co-estimation of a climatological ionospheric field as part of the CHAOS modelling approach takes into account previously unmodelled signals in the polar regions and produces geomagnetic field models that fit the magnetic input data for geomagnetically quiet conditions well. It enables the construction of high quality models of the core field while using vector field data at all latitudes and local time. Similar to the AMPS model, our approach provides estimates of the average polar ionospheric currents that are realistic in structure and able to vary in response to changes of the external driving. However, this approach is only capable to represent the long term average of the currents and not individual highly dynamic events.

One aim of this study is to answer the question of whether the co-estimation of an AMPS-type ionospheric model could allow the construction of internal field models that are less contaminated by the ionospheric field in the polar regions. By comparing the recovered SV of \modela{} and \modelref{} at the CMB in the northern polar region (see Fig.~\ref{fig:sv_panels}), we find that the co-estimation reduces the leakage of ionospheric signals into the time-dependent internal field model. However, the improvement in the recovered SV in the polar regions, which is most apparent in the zonal terms of the high spherical harmonic degrees, is relatively small and difficult to interpret due to significant ionospheric signals that remain even when co-estimating the ionospheric field model and due to the strong time averaging effect of the temporal regularisation applied on the internal field model. We find that the regularisation of the poloidal ionospheric field is important. This mostly affects the zonal or near-zonal parts of the poloidal ionospheric model as observed within an Earth-fixed frame, which are the terms that show the largest difference between \modela{} and \modelref{} in the internal field model. We acknowledge that the \modelref{} model is a somewhat extreme case since it uses vector data at high latitudes without accounting for polar ionospheric signals in any way. To test whether the ionospheric leakage into the internal magnetic field model can be further reduced in the polar regions, we derived additional test models where the internal field was estimated using the scalar component of the magnetic observations instead of vector data at polar latitudes. However, since we found these test models to be very similar to the models shown here, we preferred to present the models derived from vector data at all latitudes.

In non-polar regions we find that the estimated ionospheric field model captures the basic Solar-quiet pattern but lacks power on the dayside. Hence, there is a possibility of leakage into the time-dependent internal field due to the use of dayside data and an imperfectly modelled ionospheric field at mid and low latitudes. Comparing internal field coefficients $g_1^0$ and $g_3^0$, which are known to be the coefficients most affected by the ionospheric field at mid and low altitude, we find that the timeseries of these coefficients for \modela{}, and similarly for \modelref{}, are slightly shifted with respect to CM6, by \SIrange{2}{5}{\nT}, while the shift is smaller with respect to CHAOS-7.9. A possible future remedy could be to estimate the time-dependent internal field only from nightside data, while the ionospheric model is fit using data from all local times.

Whether the reduced ionospheric field leakage also leads to internal field models that are better resolved in time, can only be investigated by reducing the temporal regularisation that smooths the time-dependence of the internal field model, in particular affecting the high spherical harmonic degrees. For this reason we derived \modelb{}, which exhibits more power in the high degrees of the SA. However, we find that the estimated SA is now dominated at the large length-scales, approximately up to degree 8, by distinct annual oscillations. Although these annual oscillations can be filtered out of the SA estimates by using annual differences, the question of where these oscillations come from and why they are not captured by the ionospheric model remains.

A possible explanation for the leakage of annual oscillations into the internal field model is related to the parametrization of the ionospheric field, which is most likely not sufficient to account for all types of ionospheric currents and their time-dependence. In addition, it is assumed that field-aligned currents are radial, which is reasonable at high latitudes but fails at mid latitudes. Hence, it is conceivable that the annual oscillations result from seasonally varying currents at mid latitudes such as interhemispheric field-aligned currents that produce magnetic signals at satellite altitude at mid latitudes on the dayside. Accounting for these currents is challenging since it requires the estimation of poloidal and toroidal potentials of the ionospheric magnetic field within the measurement shell traced by the satellite orbits \parencite[e.g.][]{Fillion2023,Olsen1997}. Apart from ionospheric sources, other processes cannot be ruled out as the origin of the annual oscillations.

Another limitation of our models concerns the treatment of electromagnetically induced currents in Earth's interior and oceans associated with variations of the ionospheric field. Since our models are only estimated from satellite data, both the inducing and induced ionospheric fields are internal with respect to the input data. Hence, our estimated ionospheric field model also contains the induced response. In principle, through an a-posteriori analysis, the estimated ionospheric field model could be separated into induced and inducing parts using the Q-response functions \parencite[e.g.][]{Grayver2021} for a given model of the Earth's conductivity. However, this approach would not affect the quality of the core field model or resolve the ambiguity between the sources in the ionosphere and those in the core and lithosphere.

Instead of a-posteriori separating the induced and inducing parts of our estimated ionospheric field model, an alternative approach is to include the induced response during the modelling. For example, one could derive a new set of AMPS-type ionospheric field basis functions that take into account the induced counterpart via Q-response transfer functions. This would have the advantage of allowing ground-based observations to be used during the estimation of both the core and ionospheric fields, for example, using hourly mean values. For these observations, the inducing ionospheric field sources are then external, which would aid the separation of the core and ionospheric fields.

\section{Conclusions}
\label{sec:conclusions}

In this study we successfully combined the climatological approach of the AMPS model, which is suitable for modelling the ionospheric magnetic field in the polar regions, and the CHAOS modelling framework to derive models of the geomagnetic field that take into account internal and magnetospheric fields as well as the climatological aspects of the ionospheric field. We used this new approach to estimate a geomagnetic field model from satellite magnetic vector data under geomagnetic quiet conditions.

The derived model, called \modela{}, shows a good fit to the input vector data and successfully removes obvious systematic errors related to ionospheric signals in the polar regions, which were previously unaccounted for in the CHAOS modelling framework. By investigating the effect of co-estimating the ionospheric field on the internal field and its time variations in the polar regions, we find only small differences, which are most visible in the zonal terms of the high-degree spherical harmonics of the estimated SV. Importantly, high latitude non-axisymmetric SV flux features stay mostly unchanged, which adds to the evidence that they are of internal origin and therefore relevant for studies of the core flow \parencite[]{Livermore2017}.

The distinct annual oscillations in the internal field from \modelb{}, which was weakly regularised in time, could indicate that there remain ionospheric or related induced signals in the modelled internal field at low-to-mid latitudes despite the co-estimation an AMPS-type ionospheric field. This suggests shortcomings of our ionospheric field parametrization in non-polar regions, and it indicates that the noise present in time-dependent internal field models, which mostly affects the low-order spherical harmonic coefficients, is not only due to the leakage of magnetic field signals produced by high-latitude currents. Identifying the physical origin of these signals and taking them into account will be important to increase the resolution of internal field models in time.

The ambiguity between the internal field and poloidal ionospheric field models was reduced through a practical approach by regularising the time-averaged divergence-free part of the ionospheric currents, although this involves regularisation parameters that must be carefully chosen during model construction. In the future not only satellite magnetic data but also ground-based magnetic observations could be used in order to better resolve this ambiguity. But this requires treatment of the internally induced field due to the poloidal ionospheric field on the ground-based data.

%%%%%%%%%%%%%%%%%%%%%%%%%%%%%%%%%%%%%
%%  Acknowledgments
%%%%%%%%%%%%%%%%%%%%%%%%%%%%%%%%%%%%%

\begin{acknowledgments}

    We gratefully acknowledge ESA for providing access to the \swarm{} L1b data and the fully calibrated CryoSat2 magnetometer data. We wish to thank the German Aerospace Center and the Federal Ministry of Education and Research for supporting the CHAMP mission. Furthermore, we would like to thank the staff of the geomagnetic observatories and INTERMAGNET for supplying high-quality observatory data. Susan Macmillan (British Geological Survey) is gratefully acknowledged for collating checked and corrected observatory hourly mean values in the AUX OBS database. We also thank the editor and two anonymous reviewers for helpful comments and suggestions, which clarified and improved the manuscript.

    CK and CCF were funded by the European Research Council under the European Union’s Horizon 2020 research and innovation programme (grant agreement No.\ 772561). The study has been partly supported as part of \swarm{} Data Innovation and Science Cluster (\swarm{} DISC) activities, funded by the ESA contract No.\ 4000109587/13/I-NB. KL was funded by the Trond Mohn Foundation, and the AMPS model is funded by ESA through \swarm{} DISC within the reference frame of ESA contract No.\ 000109587/13/I-NB.

    CK developed the computer software used here for modelling the geomagnetic field, derived and analysed the test field models, and prepared the manuscript. CCF helped with the presentation and interpretation of the results, and assisted with the outline of the manuscript. KL helped with the interpretation of the results and provided guidance on the implementation of the part of the developed computer software that involves the modelling of the ionospheric magnetic field. NO helped with the interpretation of the results and provided the computer scripts on which parts of the developed modelling software are based. All authors read and accepted the manuscript.

\end{acknowledgments}

%%%%%%%%%%%%%%%%%%%%%%%%%%%%%%%%%%%%%
%%  Data availability
%%%%%%%%%%%%%%%%%%%%%%%%%%%%%%%%%%%%%

\begin{dataavailability}

    The data underlying this article are available in the following repositories:
    \begin{enumerate}
        \item \swarm{} and CryoSat2 data are available through ESA at \url{https://swarm-diss.eo.esa.int/#}.
        \item CHAMP data are available from \cite{Rother2019}.
        \item Ground magnetic observatory data from INTERMAGNET are available at \url{ftp://ftp.nerc-murchison.ac.uk/geomag/Swarm/AUX_OBS/hour/} or via the virtual research platform VirES \url{https://vires.services/}.
        \item $\mathit{RC}$ index is available at \url{http://www.spacecenter.dk/files/magnetic-models/RC/}.
        \item $\mathit{SML}$ index is available at \url{https://supermag.jhuapl.edu/info/}.
        \item Solar wind speed, interplanetary magnetic field and $\mathit{Kp}$ index are available through the GSFC/SPDF OMNIWeb interface at \url{https://omniweb.gsfc.nasa.gov/ow.html}.
        \item \mbox{CHAOS-7} model and its updates are available at \url{http://www.spacecenter.dk/files/magnetic-models/CHAOS-7/}.
    \end{enumerate}

    \noindent
    Files containing the estimated parameters of our models and the processed data are available at \url{https://doi.org/10.11583/DTU.24025596}.
\end{dataavailability}

%%%%%%%%%%%%%%%%%%%%%%%%%%%%%%%%%%%%%
%%  References
%%%%%%%%%%%%%%%%%%%%%%%%%%%%%%%%%%%%%

% \bibliography{references}
% \bibliographystyle{gji}

\printbibliography

%%%%%%%%%%%%%%%%%%%%%%%%%%%%%%%%%%%%%
%%  Acronyms
%%%%%%%%%%%%%%%%%%%%%%%%%%%%%%%%%%%%%

{%
\iffalse
\begin{description}
    \item[\bf CM] Comprehensive Model
    \item[\bf IMF] Interplanetary Magnetic Field
    \item[\bf CHAMP] CHAllenging Minisatellite Payload
    \item[\bf SM] Solar Magnetic
    \item[\bf GSM] Geocentric Solar Magnetic
    \item[\bf CMB] Core-Mantle Boundary
    \item[\bf QD] Quasi-Dipole
    \item[\bf MA] Modified-Apex
    \item[\bf IGRF] International Geomagnetic Reference Field
    \item[\bf MLT] Magnetic Local Time
    \item[\bf RMS] Root-Mean-Square
    \item[\bf AMPS] Average Magnetic field and Polar current System
    \item[\bf SV] Secular Variation
    \item[\bf SA] Secular Acceleration
    \item[\bf ESA] European Space Agency
    \item[\bf CRF] Common Reference Frame
    \item[\bf VFM] Vector Field Magnetometer
    \item[\bf HRN] Hornsund
    \item[\bf GRIMM] GFZ Reference Internal Magnetic Model
    \item[\bf INTERMAGNET] International Real‐time Magnetic Observatory Network
    \item[\bf EOF] Empirical Orthogonal Function
    \item[\bf PC] Principal Component
    \item[\bf PCA] Principal Component Analysis
    \item[\bf \textit{Swarm} DISC] \textit{Swarm} Data Innovation and Science Cluster
\end{description}
\fi
}%

\label{lastpage}

\end{document}